\documentclass[12pt]{article}
\newtheorem{theorem}{Theorem}
\newtheorem{definition}{Definition}

\begin{document}

\author{C. Bizdadea\thanks{%
e-mail address: bizdadea@central.ucv.ro}, S. O. Saliu\thanks{%
e-mail address: osaliu@central.ucv.ro} \\
Faculty of Physics, University of Craiova\\
13 A. I. Cuza Str., Craiova RO-1100, Romania}
\title{Lagrangian $Sp(3)$ BRST symmetry for irreducible gauge theories}
\maketitle

\begin{abstract}
The Lagrangian $Sp(3)$ BRST symmetry for irreducible gauge theories is
constructed in the framework of homological perturbation theory. The
canonical generator of this extended symmetry is shown to exist. A
gauge-fixing procedure specific to the standard antibracket-antifield
formalism, that leads to an effective action, which is invariant under all
the three differentials of the $Sp(3)$ algebra, is given.


\end{abstract}

\section{Introduction}

The crucial feature of the BRST method \cite{1}--\cite{12} is the recursive
pattern of homological perturbation theory \cite{13}--\cite{19}, which
allows one to prove the existence of the BRST symmetry itself. The machinery
of homological perturbation theory has been adapted to cover the
BRST-anti-BRST transformation in both Lagrangian and Hamiltonian
formulations \cite{20}--\cite{37}. In the context of extended BRST
symmetries, the Hamiltonian version of the $Sp(3)$ BRST symmetry has
recently been developed in \cite{38}--\cite{40}.

In this paper we show that the methods of homological perturbation theory
can be extended in order to cover the construction of the Lagrangian $Sp(3)$
BRST symmetry in the irreducible case. This can be done by triplicating each
differential appearing in the antibracket-antifield BRST formalism. We begin
with the triplication of the gauge transformations of a given irreducible
theory, which allows us to determine the correct ghost spectrum.
Consequently, we develop a proper construction of the exterior longitudinal
tricomplex to ensure that the cohomologies associated with each of the three
exterior longitudinal derivatives are isomorphic to the cohomology of the
standard exterior longitudinal derivative along the gauge orbits from the
BRST description of the initial irreducible theory. The most difficult part
is the construction of the Koszul-Tate tricomplex. Due to the fact that the
complete description of the gauge orbits obtained by triplicating the gauge
symmetries is not accompanied by the triplication of the equations of
motion, each of the three Koszul-Tate differentials contains, besides the
standard canonical part, also a noncanonical component, which acts on some
supplementary antifields (bar and tilde variables). The Koszul-Tate
tricomplex is constructed to furnish a triresolution of the algebra of
smooth functions defined on the stationary surface of field equations. The
problem of constructing the $Sp(3)$ BRST algebra is further transferred at
the level of its canonical generator, that is solution to the so-called
classical master equation of the $Sp(3)$ formalism. By means of properly
extending the homological perturbation theory, the canonical generator is
shown to exist. The final step of our treatment consists in giving a
gauge-fixing procedure that ensures the invariance of the effective action
under all three BRST symmetries that compose the Lagrangian $Sp(3)$ algebra.

The paper is organized as follows. Section 2 reviews the basic aspects of
the standard antibracket-antifield formalism. In Section 3 we present the
main ideas of the Lagrangian BRST $Sp(3)$ theory. Section 4 focuses on the
construction of the exterior longitudinal tricomplex, while Section 5
approaches the construction of the Koszul-Tate triresolution. In Section 6
we prove the existence of the canonical generator of the $Sp(3)$ theory as
solution to the extended classical master equation. The gauge-fixing
procedure is accomplished in Section 7. Section 8 exemplifies the general
procedure for abelian gauge fields. In Section 9 we present the main
conclusions of the paper.

\section{Brief review of standard antibracket-antifield formalism}

We begin with a theory described by the Lagrangian action 
\begin{equation}
S_{0}\left[ \Phi ^{\alpha _{0}}\right] =\int d^{D}x\mathcal{L}\left( \Phi
^{\alpha _{0}},\partial _{\mu _{1}}\Phi ^{\alpha _{0}},\cdots ,\partial
_{\mu _{1}\cdots \mu _{k}}\Phi ^{\alpha _{0}}\right) ,  \label{sp3.1}
\end{equation}
invariant under the (infinitesimal) gauge transformations 
\begin{equation}
\delta _{\varepsilon }\Phi ^{\alpha _{0}}=Z_{\;\;\alpha _{1}}^{\alpha
_{0}}\varepsilon ^{\alpha _{1}},  \label{sp3.2}
\end{equation}
written in De Witt's condensed notations. The fields and gauge parameters
are assumed to respectively display the Grassmann parities $\epsilon \left(
\Phi ^{\alpha _{0}}\right) \equiv \epsilon _{\alpha _{0}}$, $\epsilon \left(
\varepsilon ^{\alpha _{1}}\right) \equiv \epsilon _{\alpha _{1}}$, while
those of the gauge generators are of course $\epsilon \left( Z_{\;\;\alpha
_{1}}^{\alpha _{0}}\right) =\epsilon _{\alpha _{0}}+\epsilon _{\alpha _{1}}$%
. The action $S_{0}\left[ \Phi ^{\alpha _{0}}\right] $ is supposed to be a
local functional, so it may depend on the fields and their space-time
derivatives up to a finite order, say, $k$. In addition, the gauge
transformations (\ref{sp3.2}) are taken to be irreducible, i.e., the gauge
generators are independent\footnote{%
This means that the solutions to the equation $Z_{\;\;\alpha _{1}}^{\alpha
_{0}}f^{\alpha _{1}}\approx 0$ are trivial, $f^{\alpha _{1}}=\frac{\delta
^{R}S_{0}}{\delta \Phi ^{\alpha _{0}}}f^{\alpha _{0}\alpha _{1}}$, where the
weak equality `$\approx $' refers to the stationary surface of field
equations, $\Sigma :\frac{\delta ^{R}S_{0}}{\delta \Phi ^{\alpha _{0}}}=0$.}%
. We work in the general case of an open gauge algebra 
\begin{equation}
\frac{\delta ^{R}Z_{\;\;\alpha _{1}}^{\alpha _{0}}}{\delta \Phi ^{\beta _{0}}%
}Z_{\;\;\beta _{1}}^{\beta _{0}}-\left( -\right) ^{\epsilon _{\alpha
_{1}}\epsilon _{\beta _{1}}}\frac{\delta ^{R}Z_{\;\;\beta _{1}}^{\alpha _{0}}%
}{\delta \Phi ^{\beta _{0}}}Z_{\;\;\alpha _{1}}^{\beta _{0}}=Z_{\;\;\gamma
_{1}}^{\alpha _{0}}C_{\;\;\alpha _{1}\beta _{1}}^{\gamma _{1}}-\frac{\delta
^{R}S_{0}}{\delta \Phi ^{\beta _{0}}}M_{\alpha _{1}\beta _{1}}^{\beta
_{0}\alpha _{0}},  \label{sp3.a}
\end{equation}
where the structure functions $C_{\;\;\alpha _{1}\beta _{1}}^{\gamma _{1}}$
and $M_{\alpha _{1}\beta _{1}}^{\beta _{0}\alpha _{0}}$ may involve the
fields, possess the Grassmann parities $\epsilon \left( C_{\;\;\alpha
_{1}\beta _{1}}^{\gamma _{1}}\right) =\epsilon _{\alpha _{1}}+\epsilon
_{\beta _{1}}+\epsilon _{\gamma _{1}}$, $\epsilon \left( M_{\alpha _{1}\beta
_{1}}^{\beta _{0}\alpha _{0}}\right) =\epsilon _{\alpha _{0}}+\epsilon
_{\beta _{0}}+\epsilon _{\alpha _{1}}+\epsilon _{\beta _{1}}$, and display
the symmetry properties 
\begin{equation}
C_{\;\;\alpha _{1}\beta _{1}}^{\gamma _{1}}=-\left( -\right) ^{\epsilon
_{\alpha _{1}}\epsilon _{\beta _{1}}}C_{\;\;\beta _{1}\alpha _{1}}^{\gamma
_{1}},  \label{sp3.b}
\end{equation}
\begin{equation}
M_{\alpha _{1}\beta _{1}}^{\beta _{0}\alpha _{0}}=-\left( -\right)
^{\epsilon _{\alpha _{1}}\epsilon _{\beta _{1}}}M_{\beta _{1}\alpha
_{1}}^{\beta _{0}\alpha _{0}}=-\left( -\right) ^{\epsilon _{\alpha
_{0}}\epsilon _{\beta _{0}}}M_{\alpha _{1}\beta _{1}}^{\alpha _{0}\beta
_{0}}.  \label{sp3.c}
\end{equation}
The upper index $R$ ($L$) signifies the right (left) derivative. Taking into
account the restrictions imposed by the Jacobi identity, we are led to some
new structure functions, $D_{\;\;\alpha _{1}\beta _{1}\gamma _{1}}^{\alpha
_{0}\delta _{1}}$, defined by 
\begin{eqnarray}
&&\frac{\delta ^{R}C_{\;\;\alpha _{1}\beta _{1}}^{\delta _{1}}}{\delta \Phi
^{\alpha _{0}}}Z_{\;\;\gamma _{1}}^{\alpha _{0}}-C_{\;\;\alpha _{1}\tau
_{1}}^{\delta _{1}}C_{\;\;\beta _{1}\gamma _{1}}^{\tau _{1}}+  \nonumber \\
&&\left( -\right) ^{\epsilon _{\alpha _{1}}\left( \epsilon _{\beta
_{1}}+\epsilon _{\gamma _{1}}\right) }\left( \frac{\delta ^{R}C_{\;\;\beta
_{1}\gamma _{1}}^{\delta _{1}}}{\delta \Phi ^{\alpha _{0}}}Z_{\;\;\alpha
_{1}}^{\alpha _{0}}-C_{\;\;\beta _{1}\tau _{1}}^{\delta _{1}}C_{\;\;\gamma
_{1}\alpha _{1}}^{\tau _{1}}\right) +  \nonumber \\
&&\left( -\right) ^{\epsilon _{\gamma _{1}}\left( \epsilon _{\alpha
_{1}}+\epsilon _{\beta _{1}}\right) }\left( \frac{\delta ^{R}C_{\;\;\gamma
_{1}\alpha _{1}}^{\delta _{1}}}{\delta \Phi ^{\alpha _{0}}}Z_{\;\;\beta
_{1}}^{\alpha _{0}}-C_{\;\;\gamma _{1}\tau _{1}}^{\delta _{1}}C_{\;\;\alpha
_{1}\beta _{1}}^{\tau _{1}}\right) =  \nonumber \\
&&3\frac{\delta ^{R}S_{0}}{\delta \Phi ^{\alpha _{0}}}D_{\;\;\alpha
_{1}\beta _{1}\gamma _{1}}^{\alpha _{0}\delta _{1}}.  \label{sp3.d}
\end{eqnarray}
The procedure can be continued step by step in order to reveal the entire
higher-order tensor structure provided by the gauge theory under discussion.

At the level of the BRST formalism, the entire gauge structure of a theory
is completely captured by the BRST differential, $s$. The key point of the
antibracket-antifield formalism is represented by the construction of the
BRST differential along the general line of homological perturbation theory.
The BRST operator starts like $s=\delta +D+\cdots $, where $\delta $ is the
Koszul-Tate differential, and $D$ is the exterior longitudinal derivative
along the gauge orbits. The Koszul-Tate operator provides an homological
resolution of $C^{\infty }\left( \Sigma \right) $ (smooth functions on the
stationary surface), while $D$ takes into account the gauge invariances on $%
\Sigma $. The main feature of $s$ is its nilpotency, $s^{2}=0$. Denoting by $%
\left( ,\right) $ the antibracket, and by $S$ the canonical generator of the
Lagrangian BRST symmetry, $s\bullet =\left( \bullet ,S\right) $, the
nilpotency of $s$ is equivalent to the classical master equation 
\begin{equation}
\left( S,S\right) =0,  \label{sp3.d1}
\end{equation}
with the boundary condition 
\begin{equation}
S=S_{0}+\Phi _{\alpha _{0}}^{*}Z_{\;\;\alpha _{1}}^{\alpha _{0}}\eta
^{\alpha _{1}}+\cdots ,  \label{sp3.d2}
\end{equation}
where $\Phi _{\alpha _{0}}^{*}$ represent the antifields associated with the
original fields, and $\eta ^{\alpha _{1}}$ are the ghosts corresponding to
the gauge parameters. If we make the notation $\Phi ^{\alpha }$ for the
fields (original fields and ghosts) and $\Phi _{\alpha }^{*}$ for their
antifields, such that $\left( \Phi ^{\alpha },\Phi _{\beta }^{*}\right)
=\delta _{\beta }^{\alpha }$, we have that the Koszul-Tate differential and
the exterior longitudinal derivative along the gauge orbits can be
canonically generated through 
\begin{equation}
\delta \Phi _{\alpha }^{*}=\left. \left( \Phi _{\alpha }^{*},S\right)
\right| _{\mathrm{ghosts}=0},  \label{sp3.d3}
\end{equation}
\begin{equation}
D\Phi ^{\alpha }=\left. \left( \Phi ^{\alpha },S\right) \right| _{\mathrm{%
antifields}=0}.  \label{sp3.d4}
\end{equation}

The classical master equation is equivalent to the family of equations $%
\delta \stackrel{[k+1]}{S}=\stackrel{[k]}{D}\left[ \stackrel{[0]}{S},\cdots ,%
\stackrel{[k]}{S}\right] $, where $S=\sum\limits_{k=0}^{\infty }\stackrel{[k]%
}{S}$, $\mathrm{res}\left( \stackrel{[k]}{S}\right) =k$ and $\mathrm{gh}%
\left( \stackrel{[k]}{S}\right) =0$. The degree denoted by $\mathrm{res}$
controls the grading of the Koszul-Tate complex and is named resolution
degree, $\mathrm{res}\left( \delta \right) =-1$. The existence of the
solution to the above family is guaranteed by the acyclicity of the
Koszul-Tate differential at positive resolution degrees. With the help of
the solution to the master equation one constructs the gauge-fixed action 
\begin{equation}
S_{\psi }=S\left[ \Phi ^{\alpha },\Phi _{\alpha }^{*}=\frac{\delta ^{L}\psi 
}{\delta \Phi ^{\alpha }}\right] ,  \label{sp3.d5}
\end{equation}
where $\psi \left[ \Phi ^{\alpha }\right] $ is the gauge-fixing fermion,
which is traditionally chosen to depend only on the fields.

\section{General ideas of the Lagrangian $Sp(3)$ BRST symmetry}

Let us investigate what happens if we consider a complete redundant
description of the gauge orbits obtained by triplicating the gauge
transformations (\ref{sp3.2}) 
\begin{equation}
\delta _{\varepsilon }\Phi ^{\alpha _{0}}=Z_{\;\;\alpha _{1}}^{\alpha
_{0}}\left( \varepsilon _{1}^{\alpha _{1}}+\varepsilon _{2}^{\alpha
_{1}}+\varepsilon _{3}^{\alpha _{1}}\right) .  \label{sp3.3}
\end{equation}
Alternatively, we can write down (\ref{sp3.3}) under a matrix-like form 
\begin{equation}
\delta _{\varepsilon }\Phi ^{\alpha _{0}}=Z_{\;\;A_{1}}^{\alpha
_{0}}\varepsilon ^{A_{1}},  \label{sp3.4}
\end{equation}
where 
\begin{equation}
Z_{\;\;A_{1}}^{\alpha _{0}}=\left( Z_{\;\;\alpha _{1}}^{\alpha
_{0}},Z_{\;\;\alpha _{1}}^{\alpha _{0}},Z_{\;\;\alpha _{1}}^{\alpha
_{0}}\right) ,\;\varepsilon ^{A_{1}}=\left( 
\begin{array}{c}
\varepsilon _{1}^{\alpha _{1}} \\ 
\varepsilon _{2}^{\alpha _{1}} \\ 
\varepsilon _{3}^{\alpha _{1}}
\end{array}
\right) ,\;A_{1}=\left( \alpha _{1},\alpha _{1},\alpha _{1}\right) .
\label{sp3.5}
\end{equation}
Consequently, the gauge generators $Z_{\;\;A_{1}}^{\alpha _{0}}$ will be no
longer independent, but two-stage reducible, with the reducibility relations
and reducibility matrices respectively expressed by 
\begin{equation}
Z_{\;\;A_{1}}^{\alpha
_{0}}Z_{\;\;B_{1}}^{A_{1}}=0,\;Z_{\;\;B_{1}}^{A_{1}}Z_{\;\;\gamma
_{1}}^{B_{1}}=0,  \label{sp3.6}
\end{equation}
\begin{equation}
Z_{\;\;B_{1}}^{A_{1}}=\left( 
\begin{array}{ccc}
\mathbf{0} & \delta _{\;\;\beta _{1}}^{\alpha _{1}} & -\delta _{\;\;\beta
_{1}}^{\alpha _{1}} \\ 
-\delta _{\;\;\beta _{1}}^{\alpha _{1}} & \mathbf{0} & \delta _{\;\;\beta
_{1}}^{\alpha _{1}} \\ 
\delta _{\;\;\beta _{1}}^{\alpha _{1}} & -\delta _{\;\;\beta _{1}}^{\alpha
_{1}} & \mathbf{0}
\end{array}
\right) ,\;Z_{\;\;\gamma _{1}}^{B_{1}}=\left( 
\begin{array}{c}
-\delta _{\;\;\gamma _{1}}^{\beta _{1}} \\ 
-\delta _{\;\;\gamma _{1}}^{\beta _{1}} \\ 
-\delta _{\;\;\gamma _{1}}^{\beta _{1}}
\end{array}
\right) .  \label{sp3.7}
\end{equation}
According to the ideas of the usual antifield-BRST theory \cite{12}, we can
construct a standard BRST symmetry for the system described by the action (%
\ref{sp3.1}), subject to the second-stage reducible gauge transformations (%
\ref{sp3.3}).

Our main concern is to go deeper and see if there is possible to generate an
extended Lagrangian BRST symmetry of the type $Sp(3)$ for our gauge theory,
i.e., if we can algebraically construct three anticommuting differentials $%
\left( s_{a}\right) _{a=1,2,3}$ 
\begin{equation}
s_{a}s_{b}+s_{b}s_{a}=0,\;a,b=1,2,3.  \label{sp3.8}
\end{equation}
In other words, we investigate the construction of a differential tricomplex 
$\left( s_{1},s_{2},s_{3},\mathcal{A}\right) $, trigraded in terms of the
ghost tridegree $\mathrm{trigh}=\left( \mathrm{gh}_{1},\mathrm{gh}_{2},%
\mathrm{gh}_{3}\right) $, where 
\begin{equation}
\mathrm{trigh}\left( s_{1}\right) =\left( 1,0,0\right) ,\;\mathrm{trigh}%
\left( s_{2}\right) =\left( 0,1,0\right) ,\;\mathrm{trigh}\left(
s_{3}\right) =\left( 0,0,1\right) ,  \label{sp3.9}
\end{equation}
such that each of three differentials decomposes like 
\begin{equation}
s_{a}=\delta _{a}+D_{a}+\cdots .  \label{sp3.10}
\end{equation}
We will refer to this differential tricomplex as the $Sp(3)$ \textit{BRST
tricomplex}. The above decomposition is made according to the following
ideas:

\begin{enumerate}
\item  The three operators $\left( \delta _{a}\right) _{a=1,2,3}$ should
define a differential tricomplex $\left( \delta _{1},\delta _{2},\delta _{3},%
\mathcal{A}^{\prime }\right) $, trigraded by the resolution tridegree,
denoted by $\mathrm{trires}=\left( \mathrm{res}_{1},\mathrm{res}_{2},\mathrm{%
res}_{3}\right) $, with $\mathrm{trires}\left( \delta _{1}\right) =\left(
-1,0,0\right) $, $\mathrm{trires}\left( \delta _{2}\right) =\left(
0,-1,0\right) $, $\mathrm{trires}\left( \delta _{3}\right) =\left(
0,0,-1\right) $. Moreover, this triple complex is required to furnish a
triresolution of $C^{\infty }\left( \Sigma \right) $, where $\Sigma $ is the
stationary surface of field equations, $\Sigma :\delta ^{R}S_{0}/\delta \Phi
^{\alpha _{0}}=0$. We will call it the \textit{Koszul-Tate triresolution}.
(For a more detailed approach to triresolutions, see Appendix A.)

\item  The three operators $\left( D_{a}\right) _{a=1,2,3}$ will act on a
certain trigraded algebra $\mathcal{A}^{\prime \prime }$ such that $\left(
D_{1},D_{2},D_{3},\mathcal{A}^{\prime \prime }\right) $ is a differential
tricomplex (the \textit{exterior longitudinal tricomplex}), the
corresponding tridegree being denoted by $\mathrm{trideg}=\left( \mathrm{deg}%
_{1},\mathrm{deg}_{2},\mathrm{deg}_{3}\right) $, with $\mathrm{trideg}\left(
D_{1}\right) =\left( 1,0,0\right) $, $\mathrm{trideg}\left( D_{2}\right)
=\left( 0,1,0\right) $, $\mathrm{trideg}\left( D_{3}\right) =\left(
0,0,1\right) $. In addition, we ask that the cohomologies associated with
each $\left( D_{a}\right) _{a=1,2,3}$ are isomorphic to the cohomology of
the standard exterior longitudinal derivative along the gauge orbits from
the BRST description of the initial irreducible theory.
\end{enumerate}

The relationship between these three types of gradings is that if a
function(al) $F$ has $\mathrm{trires}\left( F\right) =\left( \mathrm{r}_{1},%
\mathrm{r}_{2},\mathrm{r}_{3}\right) $ and $\mathrm{trideg}\left( F\right)
=\left( \mathrm{d}_{1},\mathrm{d}_{2},\mathrm{d}_{3}\right) $, then its
ghost tridegree is $\mathrm{trigh}\left( F\right) =\left( \mathrm{g}_{1},%
\mathrm{g}_{2},\mathrm{g}_{3}\right) $, where $\mathrm{g}_{a}=\mathrm{d}_{a}-%
\mathrm{r}_{a}$, $a=1,2,3$. In order to implement the three symmetries $%
\left( s_{a}\right) _{a=1,2,3}$ at the Lagrangian level, we will construct
three different antibrackets, consider three extended classical master
equations, and show that there exists a common solution $S$ to these
equations, that generates the three BRST-like symmetries. In consequence, we
have to associate three antifields with each field (original fields and
ghosts). Just like in the case of the antibracket-antifield BRST-anti-BRST
symmetry, the triplication of gauge symmetries is not accompanied by a
triplication of the equations that define the stationary surface. This is
why we will need some supplementary variables (bar and tilde variables) in
the antifield sector to kill some nontrivial co-cycles in the homologies of
the Koszul-Tate operators. As a result, each of the Koszul-Tate
differentials will decompose as a sum between a canonical and a noncanonical
part, where the noncanonical operators will act only on these supplementary
antifields. Regarding the exterior longitudinal tricomplex, we will see that
a good starting point is the triplication of the gauge symmetries and the
introduction of the corresponding reducibility relations, which will allow
us to determine the correct ghost spectrum.

It is interesting to notice another aspect induced by the triplication of
the gauge symmetries. We can always define a simple differential complex
associated with a given tricomplex. For example, $\left( D,\mathcal{A}%
^{\prime \prime }\right) $, with 
\begin{equation}
D=D_{1}+D_{2}+D_{3},  \label{sp3.e}
\end{equation}
can be regarded as the simple differential complex associated with the
tricomplex $\left( D_{1},D_{2},D_{3},\mathcal{A}^{\prime \prime }\right) $
if we grade the algebra $\mathcal{A}^{\prime \prime }$ according to $\mathrm{%
deg}=\mathrm{deg}_{1}+\mathrm{deg}_{2}+\mathrm{deg}_{3}$, hence $\mathrm{deg}%
\left( D\right) =1$. The (weak) nilpotency and anticommutativity of $\left(
D_{a}\right) _{a=1,2,3}$ on $\mathcal{A}^{\prime \prime }$, $%
D_{a}D_{b}+D_{b}D_{a}\approx 0$, $a,b=1,2,3$, together with this simple
graduation, imply that 
\begin{equation}
\left( D_{a}D_{b}+D_{b}D_{a}\approx 0,\;a,b=1,2,3\right) \Leftrightarrow
D^{2}\approx 0,  \label{sp3.f}
\end{equation}
where the weak equality refers to the stationary surface. Moreover, $D$ is
nothing but the exterior longitudinal derivative associated with this new
reducible and complete description of the gauge orbits (\textit{extended
exterior longitudinal derivative}), and the cohomology of $D$ is isomorphic
to those of $\left( D_{a}\right) _{a=1,2,3}$, and therefore isomorphic to
that of the standard exterior longitudinal derivative. Along the same line, $%
\left( \delta ,\mathcal{A}^{\prime }\right) $, with 
\begin{equation}
\delta =\delta _{1}+\delta _{2}+\delta _{3},  \label{sp3.g}
\end{equation}
graded by the \textit{total resolution degree} $\mathrm{res}=\mathrm{res}%
_{1}+\mathrm{res}_{2}+\mathrm{res}_{3}$ ($\mathrm{res}\left( \delta \right)
=-1$) is precisely a simple differential complex corresponding to $\left(
\delta _{1},\delta _{2},\delta _{3},\mathcal{A}^{\prime }\right) $, and we
have that 
\begin{equation}
\left( \delta _{a}\delta _{b}+\delta _{b}\delta _{a}=0,\;a,b=1,2,3\right)
\Leftrightarrow \delta ^{2}\approx 0.  \label{sp3.h}
\end{equation}
The operator $\delta $ (\textit{extended Koszul-Tate differential}) clearly
furnishes a resolution of the algebra $C^{\infty }\left( \Sigma \right) $,
the relationship between the triresolution and resolution properties being 
\begin{equation}
\left( H_{0,0,0}\left( \delta _{a}\right) =C^{\infty }\left( \Sigma \right)
\right) \Leftrightarrow H_{0}\left( \delta \right) =C^{\infty }\left( \Sigma
\right) ,  \label{sp3.i}
\end{equation}
\begin{equation}
\left( H_{i,j,k}\left( \delta _{a}\right) =0,\;i,j,k\geq 0,\;i+j+k>0\right)
\Leftrightarrow H_{l}\left( \delta \right) =0,\;l>0,  \label{sp3.i1}
\end{equation}
where $H_{i,j,k}\left( \delta _{a}\right) $ signifies the space of elements $%
F$ with $\mathrm{trires}\left( F\right) =(i,j,k)$, that are $\delta _{a}$%
-closed modulo $\delta _{a}$-exact, and $H_{l}\left( \delta \right) $ means
the cohomological space spanned by the objects $G$ with $\mathrm{res}(G)=l$,
that are $\delta $-closed modulo $\delta $-exact.

Finally, from the $Sp(3)$ differential tricomplex $\left( s_{1},s_{2},s_{3},%
\mathcal{A}\right) $ we can construct the simple differential complex $%
\left( s,\mathcal{A}\right) $ graded according to the \textit{total ghost
number} $\mathrm{tgh}=\mathrm{gh}_{1}+\mathrm{gh}_{2}+\mathrm{gh}_{3}$,
where 
\begin{equation}
s=s_{1}+s_{2}+s_{3},  \label{sp3.j}
\end{equation}
and the\textit{\ extended BRST differential} $s$ has $\mathrm{tgh}\left(
s\right) =1$. It is clear that it splits as 
\begin{equation}
s=\delta +D+\cdots ,  \label{sp3.k}
\end{equation}
where $\delta $ and $D$ are exactly the extended Koszul-Tate differential,
respectively, the extended exterior longitudinal derivative introduced
before. The relationship between the three simple degrees is expressed by $%
\mathrm{tgh=deg}-\mathrm{res}$. The rest of the terms, denoted by `$\cdots $%
', are required in order to ensure the nilpotency of $s$, i.e., 
\begin{equation}
s^{2}=0,  \label{sp3.l}
\end{equation}
which is linked to the $Sp(3)$ algebra defining relations (\ref{sp3.8})
through 
\begin{equation}
\left( s_{a}s_{b}+s_{b}s_{a}=0,\;a,b=1,2,3\right) \Leftrightarrow s^{2}=0.
\label{sp3.m}
\end{equation}
These observations stay at the basis of the procedure we are going to
develop in order to investigate the existence of the Lagrangian BRST
symmetry of the type $Sp(3)$: instead of explicitly proving the existence of 
$\left( s_{a}\right) _{a=1,2,3}$ satisfying the properties mentioned in the
above, we will prove the existence of the associated extended BRST symmetry
and show that it splits in exactly three pieces matching the trigraduation.

\section{Exterior longitudinal tricomplex. Tricanonical structure}

We begin with the triplication of the gauge generators and gauge parameters
like in (\ref{sp3.5}), as well as with the corresponding reducibility
functions (\ref{sp3.7}). In this way, our departure point is a second-stage
reducible complete description of the gauge orbits associated with the
initial irreducible one. Following the lines of the standard BRST formalism
related to the construction of the exterior longitudinal complex in the
reducible case, we introduce the ghosts $\eta ^{A_{1}}$ for the gauge
generators $Z_{\;\;A_{1}}^{\alpha _{0}}$, the ghosts of ghosts $\pi ^{B_{1}}$
for the first-stage reducibility functions $Z_{\;\;B_{1}}^{A_{1}}$, and the
ghosts of ghosts of ghosts $\lambda ^{\gamma _{1}}$ for the second-stage
reducibility functions $Z_{\;\;\gamma _{1}}^{B_{1}}$, such that the entire
ghost spectrum is given by 
\begin{equation}
\eta ^{A_{1}}\equiv \left( \eta _{1}^{\alpha _{1}},\eta _{2}^{\alpha
_{1}},\eta _{3}^{\alpha _{1}}\right) ,\;\pi ^{B_{1}}\equiv \left( \pi
_{1}^{\beta _{1}},\pi _{2}^{\beta _{1}},\pi _{3}^{\beta _{1}}\right)
,\;\lambda ^{\gamma _{1}},  \label{sp3.11}
\end{equation}
where the Grassmann parities of the ghosts are valued like 
\begin{equation}
\epsilon \left( \eta _{a}^{\alpha _{1}}\right) =\epsilon _{\alpha
_{1}}+1,\;\epsilon \left( \pi _{a}^{\beta _{1}}\right) =\epsilon _{\beta
_{1}},\;\epsilon \left( \lambda ^{\gamma _{1}}\right) =\epsilon _{\gamma
_{1}}+1,  \label{sp3.11a}
\end{equation}
where $a=1,2,3$. The algebra $\mathcal{A}^{\prime \prime }$ will then be the
polynomial algebra generated by $\eta ^{A_{1}}$, $\pi ^{B_{1}}$ and $\lambda
^{\gamma _{1}}$, with coefficients from $C^{\infty }\left( \Sigma \right) $.
In order to properly grade this algebra, we set the obvious trigraduation 
\begin{equation}
\mathrm{trideg}\left( \eta _{1}^{\alpha _{1}}\right) =\left( 1,0,0\right) ,%
\mathrm{trideg}\left( \eta _{2}^{\alpha _{1}}\right) =\left( 0,1,0\right) ,%
\mathrm{trideg}\left( \eta _{3}^{\alpha _{1}}\right) =\left( 0,0,1\right) ,
\label{sp3.12}
\end{equation}
\begin{equation}
\mathrm{trideg}\left( \pi _{1}^{\beta _{1}}\right) =\left( 0,1,1\right) ,%
\mathrm{trideg}\left( \pi _{2}^{\beta _{1}}\right) =\left( 1,0,1\right) ,%
\mathrm{trideg}\left( \pi _{3}^{\beta _{1}}\right) =\left( 1,1,0\right) ,
\label{sp3.13}
\end{equation}
\begin{equation}
\mathrm{trideg}\left( \lambda ^{\gamma _{1}}\right) =\left( 1,1,1\right) ,\;%
\mathrm{trideg}\left( \Phi ^{\alpha _{0}}\right) =\left( 0,0,0\right) .
\label{sp3.14}
\end{equation}

By applying the rules of the standard Lagrangian BRST formalism, we can
construct the extended exterior longitudinal derivative $D$ associated with
this new reducible description of the gauge orbits. The action of $D$ on the
original fields, as well as on the generators of the polynomial algebra $%
\mathcal{A}^{\prime \prime }$, is defined by 
\begin{equation}
D\Phi ^{\alpha _{0}}=Z_{\;\;A_{1}}^{\alpha _{0}}\eta ^{A_{1}},\;D\eta
^{A_{1}}=Z_{\;\;B_{1}}^{A_{1}}\pi ^{B_{1}}+\cdots ,  \label{sp3.15a}
\end{equation}
\begin{equation}
D\pi ^{B_{1}}=Z_{\;\;\gamma _{1}}^{B_{1}}\lambda ^{\gamma _{1}}+\cdots
,\;D\lambda ^{\gamma _{1}}=\cdots ,  \label{sp3.15b}
\end{equation}
where the terms generically denoted by `$\cdots $' are such as to ensure $%
D^{2}\approx 0$, and they can be explicitly determined by using the formulas
(\ref{sp3.a}--\ref{sp3.d}) related to the gauge structure of the
investigated theory. In terms of the accompanying simple graduation, we have
that $\mathrm{deg}\left( \Phi ^{\alpha _{0}}\right) =0$, $\mathrm{deg}\left(
\eta _{a}^{\alpha _{1}}\right) =1$, $\mathrm{deg}\left( \pi _{a}^{\alpha
_{1}}\right) =2$, $a=1,2,3$, $\mathrm{deg}\left( \lambda ^{\alpha
_{1}}\right) =3$. On behalf of the relations (\ref{sp3.e}) and (\ref{sp3.15a}%
--\ref{sp3.15b}), we can write down the complete definitions of the three
exterior longitudinal derivatives on the generators from the exterior
longitudinal tricomplex in the context of the trigraduation governed by $%
\mathrm{trideg}$, under the form 
\begin{equation}
D_{a}\Phi ^{\alpha _{0}}=Z_{\;\;\alpha _{1}}^{\alpha _{0}}\eta _{a}^{\alpha
_{1}},  \label{sp3.18}
\end{equation}
\begin{equation}
D_{a}\eta _{b}^{\alpha _{1}}=\varepsilon _{abc}\pi _{c}^{\alpha _{1}}+\frac{1%
}{2}\left( -\right) ^{\epsilon _{\beta _{1}}}C_{\;\;\beta _{1}\gamma
_{1}}^{\alpha _{1}}\eta _{a}^{\gamma _{1}}\eta _{b}^{\beta _{1}},
\label{sp3.19}
\end{equation}
\begin{eqnarray}
&&D_{a}\pi _{b}^{\alpha _{1}}=-\delta _{ab}\lambda ^{\alpha _{1}}+\frac{1}{2}
\left( -\right) ^{\epsilon _{\beta _{1}}+1}C_{\;\;\beta _{1}\gamma
_{1}}^{\alpha _{1}}\eta _{a}^{\gamma _{1}}\pi _{b}^{\beta _{1}}+  \nonumber
\\
&&\frac{1}{12}\left( -\right) ^{\epsilon _{\delta _{1}}}\varepsilon
_{bcd}C_{\;\;\beta _{1}\gamma _{1}}^{\alpha _{1}}C_{\;\;\delta
_{1}\varepsilon _{1}}^{\gamma _{1}}\eta _{a}^{\varepsilon _{1}}\eta
_{c}^{\delta _{1}}\eta _{d}^{\beta _{1}},  \label{sp3.20}
\end{eqnarray}
\begin{eqnarray}
&&D_{a}\lambda ^{\alpha _{1}}=\frac{1}{2}\left( -\right) ^{\epsilon _{\beta
_{1}}}C_{\;\;\beta _{1}\gamma _{1}}^{\alpha _{1}}\eta _{a}^{\gamma
_{1}}\lambda ^{\beta _{1}}+  \nonumber \\
&&\frac{1}{12}\left( -\right) ^{\epsilon _{\delta _{1}}+1}\left(
C_{\;\;\beta _{1}\gamma _{1}}^{\alpha _{1}}C_{\;\;\delta _{1}\varepsilon
_{1}}^{\gamma _{1}}-\left( -\right) ^{\epsilon _{\beta _{1}}\left( \epsilon
_{\delta _{1}}+\epsilon _{\varepsilon _{1}}\right) }C_{\;\;\delta _{1}\gamma
_{1}}^{\alpha _{1}}C_{\;\;\varepsilon _{1}\beta _{1}}^{\gamma _{1}}\right)
\eta _{a}^{\varepsilon _{1}}\eta _{b}^{\delta _{1}}\pi _{b}^{\beta _{1}},
\label{sp3.21}
\end{eqnarray}
where $\varepsilon _{abc}$ are completely antisymmetric and constant, with $%
\varepsilon _{123}=+1$. It can be checked by direct computation that $\left(
D_{1},D_{2},D_{3},\mathcal{A}^{\prime \prime }\right) $ in the presence of
the definitions (\ref{sp3.18}--\ref{sp3.21}) indeed determines a
differential tricomplex, and, in addition, that the cohomologies associated
with each $\left( D_{a}\right) _{a=1,2,3}$ are isomorphic to the cohomology
of the standard exterior longitudinal derivative.

As we have previously noticed, we intend to define a generator that is
common to all three BRST-like symmetries $\left( s_{a}\right) _{a=1,2,3}$.
In view of this, we define three antibrackets, to be denoted by $\left(
,\right) _{a}$, with $a=1,2,3$, which further requires the introduction of
three antifields, conjugated to each field/ghost in one of the antibrackets.
In order to make this tricanonical structure compatible with the
trigraduation, and preserve the symmetry between its components, we set the
ghost tridegrees of the antibrackets as $\mathrm{trigh}\left( \left(
,\right) _{1}\right) =\left( 1,0,0\right) $, $\mathrm{trigh}\left( \left(
,\right) _{2}\right) =\left( 0,1,0\right) $, $\mathrm{trigh}\left( \left(
,\right) _{3}\right) =\left( 0,0,1\right) $. It is understood that the
antibrackets satisfy all the basic properties of the antibracket in the
standard antibracket-antifield formalism \cite{8}, \cite{12}, such as odd
Grassmann behaviour, etc. Moreover, we ask that the trigrading governing the
exterior longitudinal tricomplex does not interfere with that characteristic
to the Koszul-Tate triresolution, so we impose that $\mathrm{trideg}\left(
\delta _{a}\right) =\left( 0,0,0\right) $, $\mathrm{trires}\left(
D_{a}\right) =\left( 0,0,0\right) $, $\mathrm{trires}\left( \Phi ^{A}\right)
=\left( 0,0,0\right) $, which remains valid with respect to the
corresponding simple gradings, where $\Phi ^{A}$ is a collective notation
for all fields and ghosts 
\begin{equation}
\Phi ^{A}\equiv \left( \Phi ^{\alpha _{0}},\eta _{a}^{\alpha _{1}},\pi
_{a}^{\alpha _{1}},\lambda ^{\alpha _{1}}\right) .  \label{sp3.21a}
\end{equation}
Thus, we are led to the following antifield spectrum 
\begin{equation}
\Phi _{A}^{*(a)}\equiv \left( \Phi _{\alpha _{0}}^{*(a)},\;\eta _{b\alpha
_{1}}^{*(a)},\;\pi _{b\alpha _{1}}^{*(a)},\;\lambda _{\alpha
_{1}}^{*(a)}\right) ,\;a,b=1,2,3.  \label{sp3.22}
\end{equation}
The main features of the antifields are 
\begin{equation}
\epsilon \left( \Phi _{A}^{*(a)}\right) =\epsilon \left( \Phi ^{A}\right) +1,
\label{sp3.23}
\end{equation}
\begin{equation}
\mathrm{trires}\left( \Phi _{A}^{*(1)}\right) =\left( gh_{1}\left( \Phi
^{A}\right) +1,gh_{2}\left( \Phi ^{A}\right) ,gh_{3}\left( \Phi ^{A}\right)
\right) =-\mathrm{trigh}\left( \Phi _{A}^{*(1)}\right) ,  \label{sp3.24a}
\end{equation}
\begin{equation}
\mathrm{trires}\left( \Phi _{A}^{*(2)}\right) =\left( gh_{1}\left( \Phi
^{A}\right) ,gh_{2}\left( \Phi ^{A}\right) +1,gh_{3}\left( \Phi ^{A}\right)
\right) =-\mathrm{trigh}\left( \Phi _{A}^{*(2)}\right) ,  \label{sp3.24b}
\end{equation}
\begin{equation}
\mathrm{trires}\left( \Phi _{A}^{*(3)}\right) =\left( gh_{1}\left( \Phi
^{A}\right) ,gh_{2}\left( \Phi ^{A}\right) ,gh_{3}\left( \Phi ^{A}\right)
+1\right) =-\mathrm{trigh}\left( \Phi _{A}^{*(3)}\right) ,  \label{sp3.24c}
\end{equation}
and they result from the basic properties of the antibrackets plus the
correlation among the various trigradings. The upper index $(a)$ indicates
in which antibracket is an antifield $\Phi _{A}^{*(a)}$ conjugated to a
field $\Phi ^{A}$, such that the fundamental antibrackets read as 
\begin{equation}
\left( \Phi ^{A},\Phi _{B}^{*(a)}\right) _{b}=\delta _{B}^{A}\delta _{b}^{a}.
\label{sp3.25}
\end{equation}
For notational simplicity, we will make the convention to represent a
function(al) $F$ of resolution tridegree $\left( \mathrm{r}_{1},\mathrm{r}%
_{2},\mathrm{r}_{3}\right) $ by $\stackrel{\left[ \mathrm{r}_{1},\mathrm{r}%
_{2},\mathrm{r}_{3}\right] }{F}$ and one of ghost tridegree $\left( \mathrm{g%
}_{1},\mathrm{g}_{2},\mathrm{g}_{3}\right) $ by $\stackrel{\left( \mathrm{g}%
_{1},\mathrm{g}_{2},\mathrm{g}_{3}\right) }{F}$. Similarly, we will use the
notations $\stackrel{\left[ \mathrm{r}\right] }{F}$ and $\stackrel{\left( 
\mathrm{g}\right) }{F}$ in connection with the total resolution degree,
respectively, total ghost number. Thus, for a function(al) depending only on
the antifields, $\stackrel{\left[ \mathrm{r}_{1},\mathrm{r}_{2},\mathrm{r}%
_{3}\right] }{F}$, we can write $\stackrel{\left( -\mathrm{r}_{1},-\mathrm{r}%
_{2},-\mathrm{r}_{3}\right) }{F}$, and the same for $\mathrm{res}$ and $%
\mathrm{tgh}$. Concretely, we have that the antifields (\ref{sp3.22}) have
the following resolution tridegrees 
\begin{equation}
\stackrel{\left[ 1,0,0\right] }{\Phi }_{\alpha _{0}}^{*(1)},\stackrel{\left[
0,1,0\right] }{\Phi }_{\alpha _{0}}^{*(2)},\stackrel{\left[ 0,0,1\right] }{%
\Phi }_{\alpha _{0}}^{*(3)},\stackrel{\left[ 2,0,0\right] }{\eta }_{1\alpha
_{1}}^{*(1)},\stackrel{\left[ 1,1,0\right] }{\eta }_{1\alpha _{1}}^{*(2)},%
\stackrel{\left[ 1,0,1\right] }{\eta }_{1\alpha _{1}}^{*(3)},  \label{sp3.26}
\end{equation}
\begin{equation}
\stackrel{\left[ 1,1,0\right] }{\eta }_{2\alpha _{1}}^{*(1)},\stackrel{%
\left[ 0,2,0\right] }{\eta }_{2\alpha _{1}}^{*(2)},\stackrel{\left[
0,1,1\right] }{\eta }_{2\alpha _{1}}^{*(3)},\stackrel{\left[ 1,0,1\right] }{%
\eta }_{3\alpha _{1}}^{*(1)},\stackrel{\left[ 0,1,1\right] }{\eta }_{3\alpha
_{1}}^{*(2)},\stackrel{\left[ 0,0,2\right] }{\eta }_{3\alpha _{1}}^{*(3)},
\label{sp3.27}
\end{equation}
\begin{equation}
\stackrel{\left[ 1,1,1\right] }{\pi }_{1\alpha _{1}}^{*(1)},\stackrel{\left[
0,2,1\right] }{\pi }_{1\alpha _{1}}^{*(2)},\stackrel{\left[ 0,1,2\right] }{%
\pi }_{1\alpha _{1}}^{*(3)},\stackrel{\left[ 2,0,1\right] }{\pi }_{2\alpha
_{1}}^{*(1)},\stackrel{\left[ 1,1,1\right] }{\pi }_{2\alpha _{1}}^{*(2)},%
\stackrel{\left[ 1,0,2\right] }{\pi }_{2\alpha _{1}}^{*(3)},  \label{sp3.28}
\end{equation}
\begin{equation}
\stackrel{\left[ 2,1,0\right] }{\pi }_{3\alpha _{1}}^{*(1)},\stackrel{\left[
1,2,0\right] }{\pi }_{3\alpha _{1}}^{*(2)},\stackrel{\left[ 1,1,1\right] }{%
\pi }_{3\alpha _{1}}^{*(3)},\stackrel{\left[ 2,1,1\right] }{\lambda }%
_{\alpha _{1}}^{*(1)},\stackrel{\left[ 1,2,1\right] }{\lambda }_{\alpha
_{1}}^{*(2)},\stackrel{\left[ 1,1,2\right] }{\lambda }_{\alpha _{1}}^{*(3)},
\label{sp3.29}
\end{equation}
and hence the total resolution degrees 
\begin{equation}
\stackrel{\left[ 1\right] }{\Phi }_{\alpha _{0}}^{*(a)},\stackrel{\left[
2\right] }{\eta }_{b\alpha _{1}}^{*(a)},\stackrel{\left[ 3\right] }{\pi }%
_{b\alpha _{1}}^{*(a)},\stackrel{\left[ 4\right] }{\lambda }_{\alpha
_{1}}^{*(a)},\;a,b=1,2,3.  \label{sp3.29a}
\end{equation}

Employing the usual ideas of the BRST formalism, \textit{we ask that the
extended exterior longitudinal derivative is inferred by means of the total
antibracket }$\left( ,\right) =\left( ,\right) _{1}+\left( ,\right)
_{2}+\left( ,\right) _{3}$\textit{\ with a generator }$S$\textit{\ of total
ghost number equal to zero}. Actually, we will ask more, and take $S$ to be
of ghost tridegree $\left( 0,0,0\right) $. In this manner we are certain
that $\left( D_{a}\right) _{a=1,2,3}$ will be recovered via the antibrackets 
$\left( \left( ,\right) _{a}\right) _{a=1,2,3}$ with one and the same
generator 
\begin{equation}
D_{a}\Phi ^{A}=\left. \left( \Phi ^{A},S\right) _{a}\right| _{\Phi
_{B}^{*(b)}=0}.  \label{sp3.30}
\end{equation}
Then, from (\ref{sp3.30}) and (\ref{sp3.18}--\ref{sp3.20}), it is easy to
see that the generator $S$ begins like 
\begin{eqnarray}
&&S=S_{0}\left[ \Phi ^{\alpha _{0}}\right] +\Phi _{\alpha
_{0}}^{*(a)}Z_{\;\;\alpha _{1}}^{\alpha _{0}}\eta _{a}^{\alpha
_{1}}+\varepsilon _{abc}\eta _{b\alpha _{1}}^{*(a)}\pi _{c}^{\alpha
_{1}}-\pi _{a\alpha _{1}}^{*(a)}\lambda ^{\alpha _{1}}+\cdots =  \nonumber \\
&&\stackrel{[0]}{S}+\stackrel{[1]}{S}+\stackrel{[2]}{S}+\stackrel{[3]}{S}%
+\cdots .  \label{sp3.31}
\end{eqnarray}
The first term, $S_{0}\left[ \Phi ^{\alpha _{0}}\right] =\stackrel{[0]}{S}$
is not necessary in order to get the relations (\ref{sp3.30}), but the
reason for considering it will be clear at the level of the Koszul-Tate
triresolution. The notations $\stackrel{[n]}{S}$ are motivated by the
standard approach, and refer to a decomposition according to the total
resolution degree. All the terms involved so far with the functional $S$
satisfy the condition to have the ghost tridegree equal to $\left(
0,0,0\right) $.

\section{Koszul-Tate triresolution}

We have reached now the construction of the Koszul-Tate tricomplex.
Actually, as we want to reproduce the action of the Koszul-Tate operators on
the antifields through the antibrackets between the antifields and the
generator $S$ at all ghosts equal to zero, the form of the Koszul-Tate
operators is somehow fixed by the form of $S$. Unfortunately, this property
cannot be ensured in the framework of the Lagrangian $Sp(3)$ BRST formalism
as we require in the meantime that the Koszul-Tate tricomplex generates a
triresolution of $C^{\infty }\left( \Sigma \right) $. The incompatibility
between these two requirements is induced by the fact that the complete
description of the gauge orbits obtained by triplicating the gauge
symmetries is not accompanied by the triplication of the equations of
motion. Then, as we cannot weaken the triresolution property, we admit in
change that the Koszul-Tate operators are not entirely reproduced by means
of the antibrackets, or, in other words, we expect that these differentials
decompose in a canonical and a noncanonical part.

The boundary conditions (\ref{sp3.31}) allow us to write $\delta =\delta
_{1}+\delta _{2}+\delta _{3}$ on the polynomial algebra $\mathcal{A}^{\prime
}$ generated by the antifields $\left( \Phi _{A}^{*(a)}\right) _{a=1,2,3}$,
with coefficients from $C^{\infty }\left( I\right) $, where $I$ is the space
of all field histories. Initially, $\left( \delta _{a}\right) _{a=1,2,3}$
are defined on the generators of $\mathcal{A}^{\prime }$ through 
\begin{equation}
\delta _{a}\Phi _{\alpha _{0}}^{*(b)}=\left. \left( \Phi _{\alpha
_{0}}^{*(b)},S\right) _{a}\right| _{\mathrm{ghosts}=0}=-\delta _{ab}\frac{%
\delta ^{L}S_{0}}{\delta \Phi ^{\alpha _{0}}},  \label{sp3.32}
\end{equation}
\begin{equation}
\delta _{a}\eta _{c\alpha _{1}}^{*(b)}=\left. \left( \eta _{c\alpha
_{1}}^{*(b)},S\right) _{a}\right| _{\mathrm{ghosts}=0}=\left( -\right)
^{\epsilon _{\alpha _{1}}}\delta _{ab}\Phi _{\alpha
_{0}}^{*(c)}Z_{\;\;\alpha _{1}}^{\alpha _{0}},  \label{sp3.33}
\end{equation}
\begin{equation}
\delta _{a}\pi _{c\alpha _{1}}^{*(b)}=\left. \left( \pi _{c\alpha
_{1}}^{*(b)},S\right) _{a}\right| _{\mathrm{ghosts}=0}=\left( -\right)
^{\epsilon _{\alpha _{1}}+1}\delta _{ab}\varepsilon _{cde}\eta _{e\alpha
_{1}}^{*(d)},  \label{sp3.34}
\end{equation}
\begin{equation}
\delta _{a}\lambda _{\alpha _{1}}^{*(b)}=\left. \left( \lambda _{\alpha
_{1}}^{*(b)},S\right) _{a}\right| _{\mathrm{ghosts}=0}=\left( -\right)
^{\epsilon _{\alpha _{1}}+1}\delta _{ab}\pi _{d\alpha _{1}}^{*(d)},
\label{sp3.35}
\end{equation}
and they obviously respect the grading properties: $\mathrm{trires}\left(
\delta _{1}\right) =\left( -1,0,0\right) $, $\mathrm{trires}\left( \delta
_{2}\right) =\left( 0,-1,0\right) $, $\mathrm{trires}\left( \delta
_{3}\right) =\left( 0,0,-1\right) $, $\mathrm{res}\left( \delta \right) =%
\mathrm{res}\left( \delta _{a}\right) =-1$. However, there appear two major
problems linked to the above definitions. First of all, $\delta $ is not a
differential as $\delta ^{2}$ fails to vanish on the antifields $\pi
_{c\alpha _{1}}^{*(b)}$ and $\lambda _{\alpha _{1}}^{*(b)}$. Moreover, $%
\delta $ cannot realize a resolution of $C^{\infty }\left( \Sigma \right) $,
and $\left( \delta _{a}\right) _{a=1,2,3}$ also cannot determine a
triresolution of the same algebra, as long as there appear nontrivial
co-cycles at positive resolution degrees. Indeed, $\Phi _{A}^{*(2)}$ and $%
\Phi _{A}^{*(3)}$ are co-cycles for $\delta _{1}$, $\Phi _{A}^{*(1)}$ and $%
\Phi _{A}^{*(3)}$ for $\delta _{2}$, $\Phi _{A}^{*(1)}$ and $\Phi
_{A}^{*(2)} $ for $\delta _{3}$, while all $\Phi _{A}^{*(1)}-\Phi
_{A}^{*(2)} $, $\Phi _{A}^{*(2)}-\Phi _{A}^{*(3)}$ and $\Phi
_{A}^{*(3)}-\Phi _{A}^{*(1)} $ are nontrivial co-cycles for $\delta $.

An appropriate way to remedy both inconveniences is to add some new
variables in the antifield sector, such that all the above co-cycles become
trivial, and then modify the actions of the Koszul-Tate operators on $\pi
_{c\alpha _{1}}^{*(b)}$ and $\lambda _{\alpha _{1}}^{*(b)}$ in such a way
that they become nilpotent. Analysing the structure and the resolution
tridegrees of the nontrivial co-cycles mentioned before, it looks that we
need to introduce three such new variables (bar variables) for each triplet
of antifields 
\begin{equation}
\left( \Phi _{A}^{*(1)},\Phi _{A}^{*(2)},\Phi _{A}^{*(3)}\right) \rightarrow
\left( \bar{\Phi}_{A}^{(1)},\bar{\Phi}_{A}^{(2)},\bar{\Phi}_{A}^{(3)}\right)
,  \label{sp3.36}
\end{equation}
with the properties 
\begin{equation}
\epsilon \left( \bar{\Phi}_{A}^{(a)}\right) =\epsilon \left( \Phi
_{A}^{*(a)}\right) +1=\epsilon \left( \Phi ^{A}\right) ,\;a=1,2,3,
\label{sp3.37}
\end{equation}
\begin{equation}
\mathrm{trires}\left( \bar{\Phi}_{A}^{(1)}\right) =\left( gh_{1}\left( \Phi
^{A}\right) ,gh_{2}\left( \Phi ^{A}\right) +1,gh_{3}\left( \Phi ^{A}\right)
+1\right) ,  \label{sp3.38}
\end{equation}
\begin{equation}
\mathrm{trires}\left( \bar{\Phi}_{A}^{(2)}\right) =\left( gh_{1}\left( \Phi
^{A}\right) +1,gh_{2}\left( \Phi ^{A}\right) ,gh_{3}\left( \Phi ^{A}\right)
+1\right) ,  \label{sp3.39}
\end{equation}
\begin{equation}
\mathrm{trires}\left( \bar{\Phi}_{A}^{(3)}\right) =\left( gh_{1}\left( \Phi
^{A}\right) +1,gh_{2}\left( \Phi ^{A}\right) +1,gh_{3}\left( \Phi
^{A}\right) \right) ,  \label{sp3.40}
\end{equation}
(which further induce that $\mathrm{trigh}\left( \bar{\Phi}_{A}^{(a)}\right)
=-\mathrm{trires}\left( \bar{\Phi}_{A}^{(a)}\right) $, and $\mathrm{res}%
\left( \bar{\Phi}_{A}^{(a)}\right) =\mathrm{tgh}\left( \Phi ^{A}\right) +2=-%
\mathrm{tgh}\left( \bar{\Phi}_{A}^{(a)}\right) $), and on which $\left(
\delta _{a}\right) _{a=1,2,3}$ act like 
\begin{equation}
\delta _{a}\bar{\Phi}_{A}^{(b)}=\left( -\right) ^{\epsilon \left( \Phi
^{A}\right) }\varepsilon _{abc}\Phi _{A}^{*(c)},  \label{sp3.41}
\end{equation}
where the phase-factor in (\ref{sp3.41}) was chosen for convenience. More
precisely, the bar variables and their properties are expressed by 
\begin{equation}
\stackrel{\left[ 0,1,1\right] }{\bar{\Phi}}_{\alpha _{0}}^{(1)},\stackrel{%
\left[ 1,0,1\right] }{\bar{\Phi}}_{\alpha _{0}}^{(2)},\stackrel{\left[
1,1,0\right] }{\bar{\Phi}}_{\alpha _{0}}^{(3)},\stackrel{\left[ 1,1,1\right] 
}{\bar{\eta}}_{1\alpha _{1}}^{(1)},\stackrel{\left[ 2,0,1\right] }{\bar{\eta}%
}_{1\alpha _{1}}^{(2)},\stackrel{\left[ 2,1,0\right] }{\bar{\eta}}_{1\alpha
_{1}}^{(3)},  \label{sp3.42}
\end{equation}
\begin{equation}
\stackrel{\left[ 0,2,1\right] }{\bar{\eta}}_{2\alpha _{1}}^{(1)},\stackrel{%
\left[ 1,1,1\right] }{\bar{\eta}}_{2\alpha _{1}}^{(2)},\stackrel{\left[
1,2,0\right] }{\bar{\eta}}_{2\alpha _{1}}^{(3)},\stackrel{\left[
0,1,2\right] }{\bar{\eta}}_{3\alpha _{1}}^{(1)},\stackrel{\left[
1,0,2\right] }{\bar{\eta}}_{3\alpha _{1}}^{(2)},\stackrel{\left[
1,1,1\right] }{\bar{\eta}}_{3\alpha _{1}}^{(3)},  \label{sp3.43}
\end{equation}
\begin{equation}
\stackrel{\left[ 0,2,2\right] }{\bar{\pi}}_{1\alpha _{1}}^{(1)},\stackrel{%
\left[ 1,1,2\right] }{\bar{\pi}}_{1\alpha _{1}}^{(2)},\stackrel{\left[
1,2,1\right] }{\bar{\pi}}_{1\alpha _{1}}^{(3)},\stackrel{\left[ 1,1,2\right] 
}{\bar{\pi}}_{2\alpha _{1}}^{(1)},\stackrel{\left[ 2,0,2\right] }{\bar{\pi}}%
_{2\alpha _{1}}^{(2)},\stackrel{\left[ 2,1,1\right] }{\bar{\pi}}_{2\alpha
_{1}}^{(3)},  \label{sp3.44}
\end{equation}
\begin{equation}
\stackrel{\left[ 1,2,1\right] }{\bar{\pi}}_{3\alpha _{1}}^{(1)},\stackrel{%
\left[ 2,1,1\right] }{\bar{\pi}}_{3\alpha _{1}}^{(2)},\stackrel{\left[
2,2,0\right] }{\bar{\pi}}_{3\alpha _{1}}^{(3)},\stackrel{\left[ 1,2,2\right] 
}{\bar{\lambda}}_{\alpha _{1}}^{(1)},\stackrel{\left[ 2,1,2\right] }{\bar{%
\lambda}}_{\alpha _{1}}^{(2)},\stackrel{\left[ 2,2,1\right] }{\bar{\lambda}}%
_{\alpha _{1}}^{(3)},  \label{sp3.45}
\end{equation}
while the definitions of $\delta _{a}$ acting on them read as 
\begin{equation}
\delta _{a}\bar{\Phi}_{\alpha _{0}}^{(b)}=\left( -\right) ^{\epsilon
_{\alpha _{0}}}\varepsilon _{abc}\Phi _{\alpha _{0}}^{*(c)},\;\delta _{a}%
\bar{\eta}_{d\alpha _{1}}^{(b)}=\left( -\right) ^{\epsilon _{\alpha
_{1}}+1}\varepsilon _{abc}\eta _{d\alpha _{1}}^{*(c)},  \label{sp3.46}
\end{equation}
\begin{equation}
\delta _{a}\bar{\pi}_{d\alpha _{1}}^{(b)}=\left( -\right) ^{\epsilon
_{\alpha _{1}}}\varepsilon _{abc}\pi _{d\alpha _{1}}^{*(c)},\;\delta _{a}%
\bar{\lambda}_{\alpha _{1}}^{(b)}=\left( -\right) ^{\epsilon _{\alpha
_{1}}+1}\varepsilon _{abc}\lambda _{\alpha _{1}}^{*(c)}.  \label{sp3.47}
\end{equation}
On account of (\ref{sp3.46}--\ref{sp3.47}), we change the definitions (\ref
{sp3.34}--\ref{sp3.35}) like 
\begin{equation}
\delta _{a}\pi _{c\alpha _{1}}^{*(b)}=\left( -\right) ^{\epsilon _{\alpha
_{1}}+1}\delta _{ab}\left( \varepsilon _{cde}\eta _{e\alpha _{1}}^{*(d)}+%
\bar{\Phi}_{\alpha _{0}}^{(c)}Z_{\;\;\alpha _{1}}^{\alpha _{0}}\right) ,
\label{sp3.48}
\end{equation}
\begin{equation}
\delta _{a}\lambda _{\alpha _{1}}^{*(b)}=\left( -\right) ^{\epsilon _{\alpha
_{1}}+1}\delta _{ab}\left( \pi _{d\alpha _{1}}^{*(d)}+\bar{\eta}_{d\alpha
_{1}}^{(d)}\right) .  \label{sp3.49}
\end{equation}

At this stage, one can see that we have removed all the initial nontrivial
co-cycles from the homologies of $\delta _{a}$ and $\delta $, and
implemented the nilpotency on the antifields $\pi _{c\alpha _{1}}^{*(b)}$.
In the meantime, we have also created new nontrivial co-cycles ($\bar{\Phi}%
_{A}^{(a)}$ respectively in the homologies of $\delta _{a}$, and $\bar{\Phi}%
_{A}^{(1)}+\bar{\Phi}_{A}^{(2)}+\bar{\Phi}_{A}^{(3)}$ in that of $\delta $),
and have not yet restored the nilpotency on the generators $\lambda _{\alpha
_{1}}^{*(b)}$. In view of this, the next step will be to still enlarge the
antifield spectrum by some supplementary variables (tilde variables), one
for each triplet of bar variables 
\begin{equation}
\left( \bar{\Phi}_{A}^{(1)},\bar{\Phi}_{A}^{(2)},\bar{\Phi}_{A}^{(3)}\right)
\rightarrow \tilde{\Phi}_{A},  \label{sp3.50}
\end{equation}
with the properties 
\begin{equation}
\epsilon \left( \tilde{\Phi}_{A}\right) =\epsilon \left( \bar{\Phi}%
_{A}^{(a)}\right) +1=\epsilon \left( \Phi ^{A}\right) +1,  \label{sp3.51}
\end{equation}
\begin{equation}
\mathrm{trires}\left( \tilde{\Phi}_{A}\right) =\left( gh_{1}\left( \Phi
^{A}\right) +1,gh_{2}\left( \Phi ^{A}\right) +1,gh_{3}\left( \Phi
^{A}\right) +1\right) ,  \label{sp3.52}
\end{equation}
or, equivalently, 
\begin{equation}
\stackrel{\left[ 1,1,1\right] }{\tilde{\Phi}}_{\alpha _{0}},\stackrel{\left[
2,1,1\right] }{\tilde{\eta}}_{1\alpha _{1}},\stackrel{\left[ 1,2,1\right] }{%
\tilde{\eta}}_{2\alpha _{1}},\stackrel{\left[ 1,1,2\right] }{\tilde{\eta}}%
_{3\alpha _{1}},  \label{sp3.53}
\end{equation}
\begin{equation}
\stackrel{\left[ 1,2,2\right] }{\tilde{\pi}}_{1\alpha _{1}},\stackrel{\left[
2,1,2\right] }{\tilde{\pi}}_{2\alpha _{1}},\stackrel{\left[ 2,2,1\right] }{%
\tilde{\pi}}_{3\alpha _{1}},\stackrel{\left[ 2,2,2\right] }{\tilde{\lambda}}%
_{\alpha _{1}}.  \label{sp3.54}
\end{equation}
We define the actions of $\delta _{a}$ on the tilde variables through 
\begin{equation}
\delta _{a}\tilde{\Phi}_{A}=\left( -\right) ^{\epsilon \left( \Phi
^{A}\right) +1}\bar{\Phi}_{A}^{(a)},  \label{sp3.55}
\end{equation}
hence 
\begin{equation}
\delta _{a}\tilde{\Phi}_{\alpha _{0}}=\left( -\right) ^{\epsilon _{\alpha
_{0}}+1}\bar{\Phi}_{\alpha _{0}}^{(a)},\;\delta _{a}\tilde{\eta}_{b\alpha
_{1}}=\left( -\right) ^{\epsilon _{\alpha _{1}}}\bar{\eta}_{b\alpha
_{1}}^{(a)},  \label{sp3.56}
\end{equation}
\begin{equation}
\delta _{a}\tilde{\pi}_{b\alpha _{1}}=\left( -\right) ^{\epsilon _{\alpha
_{1}}+1}\bar{\pi}_{b\alpha _{1}}^{(a)},\;\delta _{a}\tilde{\lambda}_{\alpha
_{1}}=\left( -\right) ^{\epsilon _{\alpha _{1}}}\bar{\lambda}_{\alpha
_{1}}^{(a)},  \label{sp3.57}
\end{equation}
and adjust the definitions (\ref{sp3.49}) as 
\begin{equation}
\delta _{a}\lambda _{\alpha _{1}}^{*(b)}=\left( -\right) ^{\epsilon _{\alpha
_{1}}+1}\delta _{ab}\left( \pi _{d\alpha _{1}}^{*(d)}+\bar{\eta}_{d\alpha
_{1}}^{(d)}-\tilde{\Phi}_{\alpha _{0}}Z_{\;\;\alpha _{1}}^{\alpha
_{0}}\right) .  \label{sp3.58}
\end{equation}
In this manner, we removed all nontrivial co-cycles at positive
triresolution and resolution degrees from the homologies of $\delta _{a}$
and $\delta $, and, meanwhile, recovered the nilpotency of $\delta $.

In consequence, the total Koszul-Tate operator splits as 
\begin{equation}
\delta =\delta _{\mathrm{can}}+V,  \label{sp3.59}
\end{equation}
with $\mathrm{res}\left( \delta \right) =\mathrm{res}\left( \delta
_{can}\right) =\mathrm{res}\left( V\right) =-1$, where $\delta _{\mathrm{can}%
}$ gives the part from $\delta $ inferred via a canonical action, and the
operator $V$ acts only on the bar and tilde variables 
\begin{equation}
V=\bar{V}+\tilde{V}.  \label{sp3.60}
\end{equation}
Taking into account the relation $\delta =\delta _{1}+\delta _{2}+\delta
_{3} $ and the trigraduation, we have that 
\begin{equation}
\delta _{a}=\delta _{\mathrm{can}a}+V_{a},\;a=1,2,3,  \label{sp3.61}
\end{equation}
\begin{equation}
V_{a}=\bar{V}_{a}+\tilde{V}_{a},\;a=1,2,3,  \label{sp3.62}
\end{equation}
\begin{equation}
\delta _{\mathrm{can}a}\bullet =\left. \left( \bullet ,S\right) _{a}\right|
_{\mathrm{ghosts}=0},  \label{sp3.63}
\end{equation}
\begin{equation}
\bar{V}_{a}\bullet =\left( -\right) ^{\epsilon \left( \Phi ^{A}\right)
}\varepsilon _{abc}\Phi _{A}^{*(c)}\frac{\delta ^{R}\bullet }{\delta \bar{%
\Phi}_{A}^{(b)}},\;\tilde{V}_{a}\bullet =\left( -\right) ^{\epsilon \left(
\Phi ^{A}\right) +1}\bar{\Phi}_{A}^{(a)}\frac{\delta ^{R}\bullet }{\delta 
\tilde{\Phi}_{A}},  \label{sp3.64}
\end{equation}
hence 
\begin{equation}
\mathrm{trires}\left( V_{1}\right) =\mathrm{trires}\left( \bar{V}_{1}\right)
=\mathrm{trires}\left( \tilde{V}_{1}\right) =\left( -1,0,0\right) =-\mathrm{%
trigh}\left( V_{1}\right) ,  \label{sp3.64a}
\end{equation}
\begin{equation}
\mathrm{trires}\left( V_{2}\right) =\mathrm{trires}\left( \bar{V}_{2}\right)
=\mathrm{trires}\left( \tilde{V}_{2}\right) =\left( 0,-1,0\right) =-\mathrm{%
trigh}\left( V_{2}\right) ,  \label{sp3.64b}
\end{equation}
\begin{equation}
\mathrm{trires}\left( V_{3}\right) =\mathrm{trires}\left( \bar{V}_{3}\right)
=\mathrm{trires}\left( \tilde{V}_{3}\right) =\left( 0,0,-1\right) =-\mathrm{%
trigh}\left( V_{3}\right) .  \label{sp3.64c}
\end{equation}
Under these circumstances, it can be shown that $\left( \delta _{1},\delta
_{2},\delta _{3},\mathcal{A}^{\prime }\right) $ indeed realizes a
triresolution of $C^{\infty }\left( \Sigma \right) $, where the correct $%
\mathcal{A}^{\prime }$ is given by the polynomial algebra in the generators $%
\left( \Phi _{A}^{*(a)},\bar{\Phi}_{A}^{(a)},\tilde{\Phi}_{A}\right) $, with
coefficients from $C^{\infty }\left( I\right) $, and the actions of the
three Koszul-Tate differentials are defined with the help of the relations (%
\ref{sp3.32}--\ref{sp3.33}), (\ref{sp3.46}--\ref{sp3.48}) and (\ref{sp3.56}--%
\ref{sp3.58}). Consequently, the simple complex $\left( \delta ,\mathcal{A}%
^{\prime }\right) $ graded in terms of total resolution degree, $\mathrm{res}
$, furnishes a resolution of $C^{\infty }\left( \Sigma \right) $. In
conclusion, we had to give up the canonical action of the Koszul-Tate
differentials in favour of constructing a true triresolution of the algebra $%
C^{\infty }\left( \Sigma \right) $. Some of the most important properties of
the triresolutions, which will be used in the sequel at the proof of the
existence of $S$, are given in Appendix A.

The modification of the canonical part of $\delta _{a}$ like in (\ref{sp3.48}%
) and (\ref{sp3.58}) attracts a change in the boundary conditions for $S$ 
\begin{eqnarray}
&&S=S_{0}\left[ \Phi ^{\alpha _{0}}\right] +\Phi _{\alpha
_{0}}^{*(a)}Z_{\;\;\alpha _{1}}^{\alpha _{0}}\eta _{a}^{\alpha _{1}}+\left(
\varepsilon _{abc}\eta _{b\alpha _{1}}^{*(a)}+\bar{\Phi}_{\alpha
_{0}}^{(c)}Z_{\;\;\alpha _{1}}^{\alpha _{0}}\right) \pi _{c}^{\alpha _{1}}- 
\nonumber \\
&&\left( \pi _{a\alpha _{1}}^{*(a)}+\bar{\eta}_{a\alpha _{1}}^{(a)}-\tilde{%
\Phi}_{\alpha _{0}}Z_{\;\;\alpha _{1}}^{\alpha _{0}}\right) \lambda ^{\alpha
_{1}}+\cdots .  \label{sp3.65}
\end{eqnarray}

\section{Extended classical master equation}

Once we have determined the extended exterior longitudinal tricomplex and
realized a triresolution of $C^{\infty }\left( \Sigma \right) $, we are able
to identify the algebra $\mathcal{A}$ with the polynomial algebra 
\begin{equation}
\mathcal{A}=\mathbf{C}\left[ \Phi _{A}^{*(a)},\bar{\Phi}_{A}^{(a)},\tilde{%
\Phi}_{A}\right] \otimes C^{\infty }\left( I\right) \otimes \mathbf{C}\left[
\Phi ^{A}\right] ,  \label{sp3.66}
\end{equation}
trigraded now in terms of the ghost tridegree $\mathrm{trigh}=\left( \mathrm{%
gh}_{1},\mathrm{gh}_{2},\mathrm{gh}_{3}\right) $. We extend the actions of $%
\delta _{a}$ on the fields and ghosts by requiring that 
\begin{equation}
\delta _{a}\Phi ^{A}=0,\;a=1,2,3,  \label{sp3.66a}
\end{equation}
which ensures that $\mathrm{trigh}\left( \delta _{1}\right) =\left(
1,0,0\right) $, $\mathrm{trigh}\left( \delta _{2}\right) =\left(
0,1,0\right) $, $\mathrm{trigh}\left( \delta _{3}\right) =\left(
0,0,1\right) $ and $\mathrm{tgh}\left( \delta _{a}\right) =\mathrm{tgh}%
\left( \delta \right) =1$. Now, we have to show that there exist three
nilpotent operators $s_{a}:\mathcal{A}\rightarrow \mathcal{A}$, such that $%
\left( s_{1},s_{2},s_{3},\mathcal{A}\right) $ is a triple complex trigraded
by $\mathrm{trigh}$, and the extended BRST differential $s=s_{1}+s_{2}+s_{3}$
starts like $\delta +D+\cdots $, where the supplementary terms are chosen
such that $s^{2}=0$. Combining these results with those of the standard BRST
formalism, we are led to the construction of the generator $S$ with $\mathrm{%
tgh}\left( S\right) =0$, such that 
\begin{equation}
s\bullet =\left( \bullet ,S\right) +V\bullet ,  \label{sp3.67}
\end{equation}
where $\left( ,\right) =\left( ,\right) _{1}+\left( ,\right) _{2}+\left(
,\right) _{3}$. Demanding the nilpotency of $s$, we are led to solving the
classical master equation of the Lagrangian $Sp(3)$ formalism 
\begin{equation}
\frac{1}{2}\left( S,S\right) +VS=0,  \label{sp3.68}
\end{equation}
on account of the Jacobi identity for the antibracket and of the fact that $%
V $ behaves like a derivation with respect to the total antibracket.
Following the usual rules of homological perturbation theory, the equation (%
\ref{sp3.68}) is equivalent to the tower of equations 
\begin{equation}
\delta \stackrel{[k+1]}{S}=\stackrel{[k]}{D}\left[ \stackrel{[0]}{S},\cdots ,%
\stackrel{[k]}{S}\right] ,\;k\geq 0,  \label{sp3.69}
\end{equation}
where 
\begin{equation}
S=\sum\limits_{p=0}^{\infty }\stackrel{[p]}{S},\;\mathrm{res}\left( 
\stackrel{[p]}{S}\right) =p,\;\mathrm{tgh}\left( \stackrel{[p]}{S}\right) =0,
\label{sp3.70}
\end{equation}
while the boundary conditions on $S$ are (see (\ref{sp3.65})) 
\begin{equation}
\stackrel{\lbrack 0]}{S}=S_{0}\left[ \Phi ^{\alpha _{0}}\right] ,\;\stackrel{%
[1]}{S}=\Phi _{\alpha _{0}}^{*(a)}Z_{\;\;\alpha _{1}}^{\alpha _{0}}\eta
_{a}^{\alpha _{1}},  \label{sp3.71}
\end{equation}
\begin{equation}
\stackrel{\lbrack 2]}{S}=\left( \varepsilon _{abc}\eta _{b\alpha
_{1}}^{*(a)}+\bar{\Phi}_{\alpha _{0}}^{(c)}Z_{\;\;\alpha _{1}}^{\alpha
_{0}}\right) \pi _{c}^{\alpha _{1}}+\cdots ,  \label{sp3.72}
\end{equation}
\begin{equation}
\stackrel{\lbrack 3]}{S}=-\left( \pi _{a\alpha _{1}}^{*(a)}+\bar{\eta}%
_{a\alpha _{1}}^{(a)}-\tilde{\Phi}_{\alpha _{0}}Z_{\;\;\alpha _{1}}^{\alpha
_{0}}\right) \lambda ^{\alpha _{1}}+\cdots .  \label{sp3.73}
\end{equation}
Equations (\ref{sp3.69}) are nothing but the equations of the standard BRST
theory, so the proof of the existence of their solutions is well-known.
Indeed, it is enough to prove that $\stackrel{[k]}{D}$ is $\delta $-closed
for $k\geq 1$ in order to prove that the equations (\ref{sp3.69}) possess
solutions ($\stackrel{[0]}{S}$ and $\stackrel{[1]}{S}$ are purely boundary
terms, so they are completely expressed by (\ref{sp3.71})). On the other
hand, the $\delta $-closure of $\stackrel{[k]}{D}$ results from the Jacobi
identity for the total antibracket, so equations (\ref{sp3.69}) possess
solutions. Thus, the only matter to be dealt with remains the proof of the
fact that $s$ splits in precisely three pieces, $s=s_{1}+s_{2}+s_{3}$, of
ghost tridegrees $\left( 1,0,0\right) $, $\left( 0,1,0\right) $,
respectively, $\left( 0,0,1\right) $, or, equivalently, that $S$ is of ghost
tridegree $\left( 0,0,0\right) $.

\begin{definition}
Let $F\in \mathcal{A}$. If $F$ satisfies $\mathrm{tgh}\left( F\right) =l>0$
(respectively, $\mathrm{tgh}\left( F\right) =l\geq 0$), then $F$ is said to
be of positive ghost tridegree (respectively, nonnegative ghost tridegree)
if it can be decomposed as 
\begin{equation}
F=\sum\limits_{i+j+k=l}\stackrel{(i,j,k)}{F},  \label{sp3.74}
\end{equation}
where $i\geq 0$, $j\geq 0$ and $k\geq 0$. (We understand that by virtue of
the above notation one has $\mathrm{trigh}\stackrel{(i,j,k)}{F}=\left(
i,j,k\right) $.) Then, it is easy to see that the subset of $\mathcal{A}$
provided by the polynomials in the ghosts and antifields with coefficients
from $C^{\infty }\left( I\right) $ of positive ghost tridegree
(respectively, nonnegative ghost tridegree) is a subalgebra, to be denoted
by $\mathcal{A}_{++}$ (respectively, $\mathcal{A}_{+}$). In particular, we
have that $\mathcal{A}_{++}\subset \mathcal{A}_{+}$.
\end{definition}

Next, we prove a positivity theorem, that will be useful in the sequel.

\begin{theorem}
(positivity theorem) Let $F\in \mathcal{A}_{++}$ be such that $\mathrm{res}%
\left( F\right) =m>0$ and $\delta F=0$. Then, there exists $P\in \mathcal{A}%
_{+}$ such that $\delta P=F$.
\end{theorem}

{\textbf{Proof.}} We consider the component $\stackrel{\{i,j,k\};[p,q,r]}{F}$
of $F$ with the properties $\mathrm{trideg}\left( \stackrel{\{i,j,k\};[p,q,r]%
}{F}\right) =\left( i,j,k\right) $ and $\mathrm{trires}\left( \stackrel{%
\{i,j,k\};[p,q,r]}{F}\right) =\left( p,q,r\right) $. The assumption $\mathrm{%
res}\left( F\right) =m$ allows us to write that 
\begin{equation}
F=\sum\limits_{i,j,k}\left( \sum\limits_{p+q+r=m}\stackrel{\{i,j,k\};[p,q,r]%
}{F}\right) .  \label{sp3.75}
\end{equation}
The condition $\delta F=0$ implies that 
\begin{equation}
\delta \left( \sum\limits_{p+q+r=m}\stackrel{\{i,j,k\};[p,q,r]}{F}\right) =0,
\label{sp3.76}
\end{equation}
for every fixed triplet $\left( i,j,k\right) $, while the condition $F\in 
\mathcal{A}_{++}$ ensures that in the last sum appear only terms with $p\leq
i$, $q\leq j$ and $r\leq k$, where $i+j+k>m$. Then, it follows that we are
in the conditions of Theorem 4 (see the Appendix A), which enables us to
state that there exists an element $\stackrel{\{i,j,k\}}{P}$ with $\mathrm{%
trideg}\left( \stackrel{\{i,j,k\}}{P}\right) =\left( i,j,k\right) $ and $%
\mathrm{res}\stackrel{\{i,j,k\}}{P}=m+1$, such that 
\begin{equation}
\sum\limits_{p+q+r=m}\stackrel{\{i,j,k\};[p,q,r]}{F}=\delta \left( \stackrel{%
\{i,j,k\}}{P}\right) ,  \label{sp3.77}
\end{equation}
which can be represented like 
\begin{equation}
\stackrel{\{i,j,k\}}{P}=\sum\limits_{\bar{p}+\bar{q}+\bar{r}=m+1}\stackrel{%
\{i,j,k\};[\bar{p},\bar{q},\bar{r}]}{P},  \label{sp3.78}
\end{equation}
where in the last sum are involved only terms with $\bar{p}\leq i$, $\bar{q}%
\leq j$ and $\bar{r}\leq k$, such that the total ghost degree of every
subcomponent is $i+j+k-m-1\geq 0$ as $i+j+k>m$, which shows that $\stackrel{%
\{i,j,k\}}{P}\in \mathcal{A}_{+}$. Consequently, (\ref{sp3.75}) and (\ref
{sp3.77}) show that we have 
\begin{equation}
F=\delta P,  \label{sp3.79}
\end{equation}
with 
\begin{equation}
P=\sum\limits_{i,j,k}\stackrel{\{i,j,k\}}{P},  \label{sp3.80}
\end{equation}
where $P\in \mathcal{A}_{+}$. This proves the theorem.

In terms of the notions introduced in the above, we can equivalently
reformulate the property that $S$ is of ghost tridegree $\left( 0,0,0\right) 
$ like $S\in \mathcal{A}_{+}$. Indeed, if $\stackrel{(l,m,n)}{S}$ denotes
the component of $S$ with $\mathrm{trigh}\left( \stackrel{(l,m,n)}{S}\right)
=\left( l,m,n\right) $, then the condition $\mathrm{tgh}\left( S\right) =0$
implies that $l+m+n=0$, while $S\in \mathcal{A}_{+}$ gives that $l\geq 0$, $%
m\geq 0$, $n\geq 0$, which ensures that indeed $l=0$, $m=0$, $n=0$, hence $%
\mathrm{trigh}\left( S\right) =\left( 0,0,0\right) $. This equivalent
property is proved by the next theorem.

\begin{theorem}
Let $S$ be a solution of the classical master equation (\ref{sp3.68}). Then, 
$S$ may be chosen of nonnegative ghost tridegree, $S\in \mathcal{A}_{+}$.
\end{theorem}

{\textbf{Proof.}} We develop $S$ according to the total resolution degree
like in (\ref{sp3.70}). Suppose that $\stackrel{[p]}{S}$ is of nonnegative
ghost tridegree for $p\leq k$. Then, we have to prove that $\stackrel{[k+1]}{%
S}$ may be chosen to be of nonnegative ghost tridegree. As we have seen, the
equation (\ref{sp3.68}) is equivalent to the family (\ref{sp3.69}). (We are
only interested in the equations with $k>0$, as $\stackrel{[0]}{S}$ and $%
\stackrel{[1]}{S}$ are purely boundary terms, completely given by (\ref
{sp3.71}), and they obviously display nonnegative ghost tridegrees.) The
explicit form of $\stackrel{[k]}{D}$ is given by 
\begin{eqnarray}
&&\stackrel{[k]}{D}\left[ \stackrel{[0]}{S},\cdots ,\stackrel{[k]}{S}\right]
=-\frac{1}{2}\left( \sum\limits_{m=1}^{k}\left( \stackrel{[m]}{S},\stackrel{%
[k-m+1]}{S}\right) _{\left( \Phi ,\Phi ^{*}\right)
}+\sum\limits_{m=2}^{k}\left( \stackrel{[m]}{S},\stackrel{[k-m+2]}{S}\right)
_{\left( \eta ,\eta ^{*}\right) }+\right.  \nonumber \\
&&\left. \sum\limits_{m=3}^{k}\left( \stackrel{[m]}{S},\stackrel{[k-m+3]}{S}%
\right) _{\left( \pi ,\pi ^{*}\right) }+\sum\limits_{m=4}^{k}\left( 
\stackrel{[m]}{S},\stackrel{[k-m+4]}{S}\right) _{\left( \lambda ,\lambda
^{*}\right) }\right) ,  \label{sp3.81}
\end{eqnarray}
where $\left( ,\right) _{\left( \Phi ,\Phi ^{*}\right) }$ represents the
total antibracket constructed with respect to the fields $\Phi ^{\alpha
_{0}} $ and their antifields, $\left( ,\right) _{\left( \eta ,\eta
^{*}\right) }$ is the total antibracket involving only the ghosts $\eta
_{a}^{\alpha _{1}}$ and their antifields, and so on. On account of the
properties of the total antibracket, it is easy to see that in the right
hand-side of (\ref{sp3.81}) appear only components from $\mathcal{A}_{++}$,
hence $\stackrel{[k]}{D}\in \mathcal{A}_{++}$. Indeed, let us consider a
more general term of the type $\left( \stackrel{[p]}{S},\stackrel{[q]}{S}%
\right) $, with $p\leq k$, $q\leq k $, where $\left( ,\right) $ signifies
the total antibracket with respect to all pairs field(ghost)/antifield.
Then, as by assumption $\stackrel{[p]}{S}$ and $\stackrel{[q]}{S}$ are of
nonnegative ghost tridegree, while the possible ghost tridegrees of the
total antibracket are $\left( 1,0,0\right) $, $\left( 0,1,0\right) $ or $%
\left( 0,0,1\right) $, it follows that $\left( \stackrel{[p]}{S},\stackrel{%
[q]}{S}\right) $ is of positive ghost tridegree. Thus, the right hand-side
of the equation (\ref{sp3.69}) is of positive ghost tridegree. On the other
hand, $\delta \stackrel{[k]}{D}=0$ ($k>0$), such that we are in the
conditions of the positivity theorem. So, there exists $\stackrel{[k+1]}{S}$
such that $\delta \stackrel{[k+1]}{S}=\stackrel{[k]}{D}$, and, moreover, $%
\stackrel{[k+1]}{S}$ is of nonnegative ghost tridegree, $\stackrel{[k+1]}{S}%
\in \mathcal{A}_{+}$. This ends the proof.

As a consequence of the above theorem, we have that indeed $\mathrm{trigh}%
\left( S\right) =\left( 0,0,0\right) $, which shows that $s$ splits
precisely into three pieces $\left( s_{a}\right) _{a=1,2,3}$, of ghost
tridegrees $\left( 1,0,0\right) $, $\left( 0,1,0\right) $, respectively, $%
\left( 0,0,1\right) $. Moreover, we have that the classical master equation
of the $Sp(3)$ formalism (\ref{sp3.68}) is equivalent to three equations
corresponding to the three different antibrackets 
\begin{equation}
\frac{1}{2}\left( S,S\right) _{a}+V_{a}S=0,\;a=1,2,3.  \label{sp3.82}
\end{equation}
In this way, we have shown that there exists a generator of the Lagrangian
BRST $Sp\left( 3\right) $ symmetry, and hence this symmetry can be
constructed for any (irreducible) gauge theory. However, on the one hand,
the generator $S$ is neither $s_{1}$, nor $s_{2}$, nor $s_{3}$ invariant. On
the other hand, we need to infer a proper gauge-fixed action that can be
used in the path integral, which is $s_{a}$-invariant. In this light, we
expose in the sequel a gauge-fixing procedure that implements the $s_{a}$%
-invariances of the gauge-fixed action and presents in addition the
desirable feature of ensuring a direct equivalence with the standard
antibracket-antifield BRST formalism.

\section{Gauge-fixing procedure}

We begin by restoring an anticanonical structure for all the variables (and,
in particular, for the bar and tilde ones) for bringing the classical master
equation of the BRST $Sp(3)$ formalism to a more familiar form. In view of
this, we focus for the moment on a single antibracket, for example on the
first one, and forget about the other two, $\left( ,\right) _{2}$ and $%
\left( ,\right) _{3}$. In order to preserve the tricanonical structure, we
cannot declare the existing variables, excepting $\Phi ^{A}$ and $\Phi
_{A}^{*(1)}$, conjugated in the first antibracket. This is why we need to
extend the algebra of the BRST $Sp(3)$ tricomplex $\mathcal{A}$ by adding
the variables 
\begin{equation}
\left( \rho _{2}^{A},\rho _{3}^{A},\kappa _{1}^{A},\mu _{2}^{A},\mu
_{3}^{A},\nu _{1}^{A}\right) ,  \label{sp3.100a}
\end{equation}
with the properties 
\begin{equation}
\epsilon \left( \rho _{2}^{A}\right) =\epsilon \left( \rho _{3}^{A}\right)
=\epsilon \left( \nu _{1}^{A}\right) =\epsilon \left( \Phi ^{A}\right) ,
\label{sp3.100}
\end{equation}
\begin{equation}
\epsilon \left( \kappa _{1}^{A}\right) =\epsilon \left( \mu _{2}^{A}\right)
=\epsilon \left( \mu _{3}^{A}\right) =\epsilon \left( \Phi ^{A}\right) +1,
\label{sp3.101}
\end{equation}
\begin{equation}
\mathrm{trigh}\left( \rho _{2}^{A}\right) =\left( \mathrm{gh}_{1}\Phi ^{A}-1,%
\mathrm{gh}_{2}\Phi ^{A},\mathrm{gh}_{3}\Phi ^{A}+1\right) ,  \label{sp3.102}
\end{equation}
\begin{equation}
\mathrm{trigh}\left( \rho _{3}^{A}\right) =\left( \mathrm{gh}_{1}\Phi ^{A}-1,%
\mathrm{gh}_{2}\Phi ^{A}+1,\mathrm{gh}_{3}\Phi ^{A}\right) ,  \label{sp3.103}
\end{equation}
\begin{equation}
\mathrm{trigh}\left( \nu _{1}^{A}\right) =\left( \mathrm{gh}_{1}\Phi ^{A},%
\mathrm{gh}_{2}\Phi ^{A}+1,\mathrm{gh}_{3}\Phi ^{A}+1\right) ,
\label{sp3.104}
\end{equation}
\begin{equation}
\mathrm{trigh}\left( \kappa _{1}^{A}\right) =\left( \mathrm{gh}_{1}\Phi
^{A}-1,\mathrm{gh}_{2}\Phi ^{A}+1,\mathrm{gh}_{3}\Phi ^{A}+1\right) ,
\label{sp3.105}
\end{equation}
\begin{equation}
\mathrm{trigh}\left( \mu _{2}^{A}\right) =\left( \mathrm{gh}_{1}\Phi ^{A},%
\mathrm{gh}_{2}\Phi ^{A}+1,\mathrm{gh}_{3}\Phi ^{A}\right) ,  \label{sp3.106}
\end{equation}
\begin{equation}
\mathrm{trigh}\left( \mu _{3}^{A}\right) =\left( \mathrm{gh}_{1}\Phi ^{A},%
\mathrm{gh}_{2}\Phi ^{A},\mathrm{gh}_{3}\Phi ^{A}+1\right) ,  \label{sp3.107}
\end{equation}
\begin{equation}
\left( \Phi _{A}^{*(3)},\rho _{2}^{B}\right) _{1}=\delta _{A}^{B},\;\left(
\rho _{3}^{A},\Phi _{B}^{*(2)}\right) _{1}=\delta _{B}^{A},\;\left( \bar{\Phi%
}_{A}^{(1)},\kappa _{1}^{B}\right) _{1}=\delta _{A}^{B},  \label{sp3.108}
\end{equation}
\begin{equation}
\left( \mu _{3}^{A},\bar{\Phi}_{B}^{(2)}\right) _{1}=\delta
_{B}^{A},\;\left( \bar{\Phi}_{A}^{(3)},\mu _{2}^{B}\right) _{1}=\delta
_{A}^{B},\;\left( \nu _{1}^{A},\tilde{\Phi}_{B}\right) _{1}=\delta _{B}^{A}.
\label{sp3.109}
\end{equation}
So, if $S$ is solution to the equation (\ref{sp3.68}), then 
\begin{equation}
S_{1}=S+\Phi _{A}^{*(2)}\mu _{2}^{A}+\Phi _{A}^{*(3)}\mu _{3}^{A}+\bar{\Phi}%
_{A}^{(1)}\nu _{1}^{A},  \label{sp3.109a}
\end{equation}
satisfies the equation 
\begin{equation}
\left( S_{1},S_{1}\right) _{1}=0,  \label{sp3.109b}
\end{equation}
which is nothing but the standard classical master equation in the first
antibracket. As a consequence, we can apply the gauge-fixing procedure from
the standard antibracket-antifield approach. In view of this, we have to
choose a certain fermionic functional $\psi _{1}$ that depends on half of
the variables from the enlarged BRST tricomplex, and eliminate the other
half with its help. We notice that, according to the fundamental
antibrackets (\ref{sp3.108}--\ref{sp3.109}) and $\left( \Phi ^{A},\Phi
_{B}^{*(1)}\right) _{1}=\delta _{B}^{A}$, the variables $\Phi ^{A}$, $\Phi
_{A}^{*(3)}$, $\rho _{3}^{A}$, $\bar{\Phi}_{A}^{(1)}$, $\mu _{3}^{A}$, $\bar{%
\Phi}_{A}^{(3)}$ and $\nu _{1}^{A}$ are regarded as `fields', while $\Phi
_{A}^{*(1)}$, $\rho _{2}^{A}$, $\Phi _{A}^{*(2)}$, $\kappa _{1}^{A}$, $\bar{%
\Phi}_{A}^{(2)}$, $\mu _{2}^{A}$ and $\tilde{\Phi}_{A}$ are viewed like
their corresponding `antifields'. We choose to eliminate the variables $\Phi
_{A}^{*(1)}$, $\rho _{2}^{A}$, $\rho _{3}^{A}$, $\kappa _{1}^{A}$, $\bar{\Phi%
}_{A}^{(2)}$, $\bar{\Phi}_{A}^{(3)}$ and $\tilde{\Phi}_{A}$, and, moreover,
to take the fermion $\psi _{1}$ such as to implement the gauge-fixing
conditions 
\begin{equation}
\rho _{2}^{A}=\rho _{3}^{A}=\kappa _{1}^{A}=0.  \label{sp3.110}
\end{equation}
Then, we have to take $\psi _{1}$ not to depend on the variables conjugated
to those appearing in (\ref{sp3.110}), namely, $\Phi _{A}^{*(3)}$, $\Phi
_{A}^{*(2)}$ and $\bar{\Phi}_{A}^{(1)}$, but only on the remaining ones,
i.e., $\Phi ^{A}$, $\mu _{2}^{A}$, $\mu _{3}^{A}$ and $\nu _{1}^{A}$. (The
variables are eliminated from the theory with the aid of the well-known
relations $\mathrm{antifield}=\frac{\delta ^{L}\psi _{1}}{\delta \left( 
\mathrm{field}\right) }$,\ $\mathrm{field}=-\frac{\delta ^{L}\psi _{1}}{%
\delta \left( \mathrm{antifield}\right) }$, where it is understood that the
`field' is conjugated to the corresponding `antifield' in the first
antibracket.) In view of this, we take $\psi _{1}$ of the form 
\begin{equation}
\psi _{1}=\left( -\right) ^{1+\epsilon \left( \Phi ^{B}\right) \left(
\epsilon \left( \Phi ^{A}\right) +1\right) }\frac{\delta ^{2L}\psi \left[
\Phi ^{A}\right] }{\delta \Phi ^{A}\delta \Phi ^{B}}\mu _{3}^{A}\mu _{2}^{B}+%
\frac{\delta ^{L}\psi \left[ \Phi ^{A}\right] }{\delta \Phi ^{A}}\nu
_{1}^{A},  \label{sp3.111}
\end{equation}
where $\psi \left[ \Phi ^{A}\right] $ is an arbitrary fermionic functional
involving only the fields and ghosts. From (\ref{sp3.111}) we find the
gauge-fixing conditions (\ref{sp3.110}), as well as 
\begin{equation}
\Phi _{A}^{*(1)}=\left( -\right) ^{1+\epsilon \left( \Phi ^{C}\right) \left(
\epsilon \left( \Phi ^{B}\right) +1\right) }\frac{\delta ^{3L}\psi }{\delta
\Phi ^{A}\delta \Phi ^{B}\delta \Phi ^{C}}\mu _{3}^{B}\mu _{2}^{C}+\frac{%
\delta ^{2L}\psi }{\delta \Phi ^{A}\delta \Phi ^{B}}\nu _{1}^{B}\equiv
f_{A}\left( \Phi ,\mu ,\nu \right) ,  \label{sp3.112}
\end{equation}
\begin{equation}
\bar{\Phi}_{A}^{(2)}=\left( -\right) ^{\epsilon \left( \Phi ^{A}\right) }%
\frac{\delta ^{2L}\psi }{\delta \Phi ^{A}\delta \Phi ^{B}}\mu _{2}^{B}\equiv
g_{2A}\left( \Phi ,\mu \right) ,  \label{sp3.113}
\end{equation}
\begin{equation}
\bar{\Phi}_{A}^{(3)}=\left( -\right) ^{\epsilon \left( \Phi ^{A}\right) }%
\frac{\delta ^{2L}\psi }{\delta \Phi ^{A}\delta \Phi ^{B}}\mu _{3}^{B}\equiv
g_{3A}\left( \Phi ,\mu \right) ,  \label{sp3.114}
\end{equation}
\begin{equation}
\tilde{\Phi}_{A}=\frac{\delta ^{L}\psi }{\delta \Phi ^{A}}\equiv h_{A}\left(
\Phi \right) ,  \label{sp3.115}
\end{equation}
such that the gauge-fixed action will be given by 
\begin{eqnarray}
&&S_{1\psi }\left[ \Phi ^{A},\Phi _{A}^{*(2)},\Phi _{A}^{*(3)},\bar{\Phi}%
_{A}^{(1)},\mu _{2}^{A},\mu _{3}^{A},\nu _{1}^{A}\right] =  \nonumber \\
&&S\left[ \Phi ^{A},\Phi _{A}^{*(1)}=f_{A},\Phi _{A}^{*(2)},\Phi _{A}^{*(3)},%
\bar{\Phi}_{A}^{(1)},\bar{\Phi}_{A}^{(2)}=g_{2A},\bar{\Phi}_{A}^{(3)}=g_{3A},%
\tilde{\Phi}_{A}=h_{A}\right] +  \nonumber \\
&&\Phi _{A}^{*(2)}\mu _{2}^{A}+\Phi _{A}^{*(3)}\mu _{3}^{A}+\bar{\Phi}%
_{A}^{(1)}\nu _{1}^{A}.  \label{sp3.116}
\end{eqnarray}
In consequence, after applying the gauge-fixing procedure, we have reached
the gauge-fixed action (\ref{sp3.116}). Moreover, it can be simply shown
that it leads to an effective action which is $s_{a}$-invariant, and that
can be further used in the path integral. Although correct, the gauge-fixing
procedure developed so far is not completely satisfactory in a sense that
will be made clear below. Let us consider the simple case of an abelian
gauge algebra with gauge generators $Z_{\;\;\alpha _{1}}^{\alpha _{0}}$
independent of the fields, such that $\epsilon _{\alpha _{0}}=\epsilon
_{\alpha _{1}}$. In this case, the solution to the master equation (\ref
{sp3.109b}) is completely given by (\ref{sp3.109a}), where $S$ is entirely
expressed by the boundary conditions (\ref{sp3.65}). We take a gauge-fixing
fermion $\psi $ that is quadratic in the fields $\Phi ^{A}$. From (\ref
{sp3.112}--\ref{sp3.115}) it follows that the functions $f_{A}$, $g_{2A}$
and $g_{3A}$ will not involve the $\Phi ^{A}$'s, while $h_{A}$ are linear in 
$\Phi ^{A}$. Then, $h_{A}$ induces in the gauge-fixed action the term $%
h_{\alpha _{1}}\left( \Phi ^{A}\right) \lambda ^{\alpha _{1}}$. Because $%
\epsilon \left( h_{\alpha _{1}}\left( \Phi ^{A}\right) \right) =\epsilon
_{\alpha _{1}}+1=\epsilon _{\alpha _{0}}+1$, and $h_{\alpha _{1}}$ are
linear in $\Phi ^{A}$, it results that $h_{\alpha _{1}}$ cannot depend on
the original fields, $\Phi ^{\alpha _{0}}$ (as $\epsilon \left( \Phi
^{\alpha _{0}}\right) =\epsilon _{\alpha _{0}}$). This means that at the
level of the gauge-fixed action (\ref{sp3.116}) the only dependence of the
original fields appears in the original Lagrangian action itself, hence the
gauge-fixed action is degenerate (due to the gauge invariances (\ref{sp3.2}%
)). Thus, in order to infer a correct gauge-fixed action we must take $\psi $
at least cubic in the $\Phi ^{A}$'s. Now, we assume that $\psi $ is cubic in
the $\Phi ^{A}$'s. In this situation, from (\ref{sp3.112}--\ref{sp3.115}) we
observe that the functions $f_{A}$, $g_{2A}$, $g_{3A}$ and $h_{A}$ will
depend in general on $\Phi ^{\alpha _{0}}$, being quadratic in the variables
they involve, such that they will induce within the gauge-fixed action (\ref
{sp3.116}) a dependence of $\Phi ^{\alpha _{0}}$ that is manifested through
interaction terms that are cubic in the various variables. In conclusion,
the gauge-fixed action cannot contain a gauge-fixing term of the type $%
b^{\alpha _{1}}f_{\alpha _{1}}\left( \Phi ^{\alpha _{0}}\right) $, where $%
f_{\alpha _{1}}\left( \Phi ^{\alpha _{0}}\right) =0$ signify the gauge
conditions on the original fields. This fact represents an inconvenient
because the presence of such a term is necessary for a proper relationship
with the standard antibracket-antifield formalism.

In order to surpass this inconvenient, we proceed as follows. We introduce
the purely gauge fields $\varphi ^{\alpha _{1}}$ (that do not enter the
original action) with the gauge invariances $\delta _{\xi }\varphi ^{\alpha
_{1}}=\xi ^{\alpha _{1}}$, where the gauge parameters $\xi ^{\alpha _{1}}$
have statistics opposite to that of the gauge conditions $f_{\alpha
_{1}}\left( \Phi ^{\alpha _{0}}\right) $, hence we have 
\begin{equation}
\epsilon \left( \varphi ^{\alpha _{1}}\right) =\epsilon \left( \xi ^{\alpha
_{1}}\right) =\epsilon \left( f_{\alpha _{1}}\right) +1.  \label{sp3.119}
\end{equation}
According to the general theory exposed in section 4, we introduce the
additional ghost sector 
\begin{equation}
\left( \stackrel{(1,0,0)}{C}_{1}^{\alpha _{1}},\stackrel{(0,1,0)}{C}%
_{2}^{\alpha _{1}},\stackrel{(0,0,1)}{C}_{3}^{\alpha _{1}},\stackrel{(0,1,1)%
}{p}_{1}^{\alpha _{1}},\stackrel{(1,0,1)}{p}_{2}^{\alpha _{1}},\stackrel{%
(1,1,0)}{p}_{3}^{\alpha _{1}},\stackrel{(1,1,1)}{l}^{\alpha _{1}}\right) ,
\label{sp3.120}
\end{equation}
displaying the Grassmann parities 
\begin{equation}
\epsilon \left( C_{a}^{\alpha _{1}}\right) =\epsilon \left( l^{\alpha
_{1}}\right) =\epsilon \left( f_{\alpha _{1}}\right) ,\;\epsilon \left(
p_{a}^{\alpha _{1}}\right) =\epsilon \left( f_{\alpha _{1}}\right)
+1,\;a=1,2,3.  \label{sp3.121}
\end{equation}
For notational simplicity, we make the collective notation 
\begin{equation}
\varphi ^{I}=\left( \varphi ^{\alpha _{1}},C_{a}^{\alpha _{1}},p_{a}^{\alpha
_{1}},l^{\alpha _{1}}\right) .  \label{sp3.122}
\end{equation}
Thus, as explained in sections 4 and 5, the antifield spectrum will contain
the variables 
\begin{equation}
\left( \varphi _{I}^{*(a)},\bar{\varphi}_{I}^{(a)},\tilde{\varphi}%
_{I}\right) ,\;a=1,2,3,  \label{sp3.123}
\end{equation}
whose properties result from the general formulas (\ref{sp3.23}--\ref
{sp3.24c}), (\ref{sp3.37}--\ref{sp3.40}) and (\ref{sp3.51}--\ref{sp3.52})
adapted to the additional field/ghost spectrum. As the sector corresponding
to the new fields does not interfere in any point with the original one, the
solution to the master equation of the BRST $Sp(3)$ formalism associated
with the overall gauge theory will be 
\begin{equation}
\bar{S}=S+\varphi _{\alpha _{1}}^{*(a)}C_{a}^{\alpha _{1}}+\left(
\varepsilon _{abc}C_{b\alpha _{1}}^{*(a)}+\bar{\varphi}_{\alpha
_{1}}^{(c)}\right) p_{c}^{\alpha _{1}}-\left( p_{a\alpha _{1}}^{*(a)}+\bar{C}%
_{a\alpha _{1}}^{(a)}-\tilde{\varphi}_{\alpha _{1}}\right) l^{\alpha _{1}}.
\label{sp3.132}
\end{equation}
Now, we reprise the gauge-fixing procedure exposed in the above, but with
respect to the larger gauge theory. We give up the second and third
antibrackets, and introduce the variables $\left(
r_{2}^{I},r_{3}^{I},k_{1}^{I},m_{2}^{I},m_{3}^{I},n_{1}^{I}\right) $ with
the properties respectively of the type (\ref{sp3.100}--\ref{sp3.109}).
Subsequently, we pass to the functional 
\begin{equation}
\bar{S}_{1}=\bar{S}+\Phi _{A}^{*(2)}\mu _{2}^{A}+\Phi _{A}^{*(3)}\mu
_{3}^{A}+\bar{\Phi}_{A}^{(1)}\nu _{1}^{A}+\varphi
_{I}^{*(2)}m_{2}^{I}+\varphi _{I}^{*(3)}m_{3}^{I}+\bar{\varphi}%
_{I}^{(1)}n_{1}^{I},  \label{sp3.133}
\end{equation}
which is of course solution to the standard classical master equation in the
first antibracket, $\left( \bar{S}_{1},\bar{S}_{1}\right) _{1}=0$. Now, we
have to choose a fermionic functional $\bar{\psi}_{1}$, with the help of
which we eliminate half of the variables in favour of the other half. We act
like before, and eliminate, besides $\Phi _{A}^{*(1)}$, $\rho _{2}^{A}$, $%
\rho _{3}^{A}$, $\kappa _{1}^{A}$, $\bar{\Phi}_{A}^{(2)}$, $\bar{\Phi}%
_{A}^{(3)}$ and $\tilde{\Phi}_{A}$, also the variables $\varphi _{I}^{*(1)}$%
, $r_{2}^{I}$, $r_{3}^{I}$, $k_{1}^{I}$, $\bar{\varphi}_{I}^{(2)}$, $\bar{%
\varphi}_{I}^{(3)}$ and $\tilde{\varphi}_{I}$ via the gauge-fixing
conditions (\ref{sp3.110}) together with 
\begin{equation}
r_{2}^{I}=r_{3}^{I}=k_{1}^{I}=0.  \label{sp3.134}
\end{equation}
This can be done through 
\begin{equation}
\bar{\psi}_{1}=\left( -\right) ^{1+\epsilon \left( \Phi ^{\Lambda }\right)
\left( \epsilon \left( \Phi ^{\Delta }\right) +1\right) }\frac{\delta ^{2L}%
\bar{\psi}^{\prime }\left[ \Phi ^{\Delta }\right] }{\delta \Phi ^{\Delta
}\delta \Phi ^{\Lambda }}\mu _{3}^{\Delta }\mu _{2}^{\Lambda }+\frac{\delta
^{L}\bar{\psi}^{\prime }\left[ \Phi ^{\Delta }\right] }{\delta \Phi ^{\Delta
}}\nu _{1}^{\Delta },  \label{sp3.135}
\end{equation}
where the fermionic functional $\bar{\psi}^{\prime }$ depends only on $\Phi
^{\Delta }=\left( \Phi ^{A},\varphi ^{I}\right) $. Also, we employed the
notations $\mu _{2}^{\Delta }=\left( \mu _{2}^{A},m_{2}^{I}\right) $, $\mu
_{3}^{\Delta }=\left( \mu _{3}^{A},m_{3}^{I}\right) $, $\nu _{1}^{\Delta
}=\left( \nu _{1}^{A},n_{1}^{I}\right) $. In order to implement some linear
gauge conditions, we take the functional $\bar{\psi}^{\prime }$ of the
special form 
\begin{equation}
\bar{\psi}^{\prime }=\bar{\psi}\left[ \Phi ^{\Delta }\right] +\varphi
^{\alpha _{1}}f_{\alpha _{1}}\left( \Phi ^{\alpha _{0}}\right) ,
\label{sp3.136}
\end{equation}
where $f_{\alpha _{1}}\left( \Phi ^{\alpha _{0}}\right) $ is linear in the
fields $\Phi ^{\alpha _{0}}$. The elimination process gives, at the level of
the larger gauge theory, some relations of the type (\ref{sp3.112}--\ref
{sp3.115}), with $A$, $B$, $C$ replaced by $\Delta $, $\Delta ^{\prime }$, $%
\Delta ^{\prime \prime }$, and $\psi $ by $\bar{\psi}^{\prime }$. We remark
that the elimination of the variables connected with the fields $\Phi
^{\alpha _{0}}$ and $\varphi ^{\alpha _{1}}$ produces the terms 
\begin{equation}
\Phi _{\alpha _{0}}^{*(1)}=N_{\alpha _{1}\alpha _{0}}n_{1}^{\alpha
_{1}}+\cdots ,\;\varphi _{\alpha _{1}}^{*(1)}=\left( -\right) ^{\epsilon
_{\alpha _{0}}\left( \epsilon \left( f_{\alpha _{1}}\right) +1\right)
}N_{\alpha _{1}\alpha _{0}}\nu _{1}^{\alpha _{0}}+\cdots ,  \label{sp3.137}
\end{equation}
\begin{equation}
\bar{\Phi}_{\alpha _{0}}^{(2)}=\left( -\right) ^{\epsilon _{\alpha
_{0}}}N_{\alpha _{1}\alpha _{0}}m_{2}^{\alpha _{1}}+\cdots ,\;\bar{\varphi}%
_{\alpha _{1}}^{(2)}=\left( -\right) ^{\left( \epsilon _{\alpha
_{0}}+1\right) \left( \epsilon \left( f_{\alpha _{1}}\right) +1\right)
}N_{\alpha _{1}\alpha _{0}}\mu _{2}^{\alpha _{0}}+\cdots ,  \label{sp3.138}
\end{equation}
\begin{equation}
\bar{\Phi}_{\alpha _{0}}^{(3)}=\left( -\right) ^{\epsilon _{\alpha
_{0}}}N_{\alpha _{1}\alpha _{0}}m_{3}^{\alpha _{1}}+\cdots ,\;\bar{\varphi}%
_{\alpha _{1}}^{(3)}=\left( -\right) ^{\left( \epsilon _{\alpha
_{0}}+1\right) \left( \epsilon \left( f_{\alpha _{1}}\right) +1\right)
}N_{\alpha _{1}\alpha _{0}}\mu _{3}^{\alpha _{0}}+\cdots ,  \label{sp3.139}
\end{equation}
\begin{equation}
\tilde{\Phi}_{\alpha _{0}}=N_{\alpha _{1}\alpha _{0}}\varphi ^{\alpha
_{1}}+\cdots ,\;\tilde{\varphi}_{\alpha _{1}}=f_{\alpha _{1}}\left( \Phi
^{\alpha _{0}}\right) +\cdots ,  \label{sp3.140}
\end{equation}
where $N_{\alpha _{1}\alpha _{0}}=\frac{\delta ^{L}f_{\alpha _{1}}}{\delta
\Phi ^{\alpha _{0}}}$.

The gauge-fixed action $\bar{S}_{1\bar{\psi}^{\prime }}\left[ \Phi ^{\Delta
},\Phi _{\Delta }^{*(2)},\Phi _{\Delta }^{*(3)},\bar{\Phi}_{\Delta
}^{(1)},\mu _{2}^{\Delta },\mu _{3}^{\Delta },\nu _{1}^{\Delta }\right] $
will be constructed from (\ref{sp3.133}) where we eliminate some of the
`fields' and `antifields' with the help of the expressions resulting from $%
\bar{\psi}^{\prime }$, such that the path integral can be written as 
\begin{equation}
Z_{\bar{\psi}^{\prime }}=\int \mathcal{D}\Phi ^{\Delta }\mathcal{D}\Phi
_{\Delta }^{*(2)}\mathcal{D}\Phi _{\Delta }^{*(3)}\mathcal{D}\bar{\Phi}%
_{\Delta }^{(1)}\mathcal{D}\mu _{2}^{\Delta }\mathcal{D}\mu _{3}^{\Delta }%
\mathcal{D}\nu _{1}^{\Delta }\exp \left( i\bar{S}_{1\bar{\psi}^{\prime
}}\right) ,  \label{sp3.141}
\end{equation}
being understood that $\Phi _{\Delta }^{*(a)}=\left( \Phi
_{A}^{*(a)},\varphi _{I}^{*(a)}\right) $, $\bar{\Phi}_{\Delta }^{(a)}=\left( 
\bar{\Phi}_{A}^{(a)},\bar{\varphi}_{I}^{(a)}\right) $. If we take $\bar{\psi}
$ not to depend on $\varphi ^{\alpha _{1}}$, $C_{a}^{\alpha _{1}}$ and $%
p_{a}^{\alpha _{1}}$, the latter relations in (\ref{sp3.140}) yield the term 
$f_{\alpha _{1}}\left( \Phi ^{\alpha _{0}}\right) l^{\alpha _{1}}$ in the
gauge-fixed action, which represents the standard gauge-fixing term. The
gauge-fixing procedure developed in relation with the larger theory is more
flexible. Excepting the fact that we can obtain a standard gauge-fixing
term, it is no longer necessary that $\bar{\psi}$ is cubic in the $\Phi
^{\Delta }$'s as the term that fixes the gauge invariances of the original
fields comes now from the second piece in (\ref{sp3.136}), so $\bar{\psi}$
may be taken to be quadratic (see also the example).

Now, we define the effective action. This is obtained by enforcing the
gauge-fixing conditions directly in the solution $\bar{S}_{1}$ of the
classical master equation in the first antibracket with the help of some
Lagrange multipliers 
\begin{eqnarray}
&&S_{\mathrm{eff}}=\bar{S}_{1}+\left( \Phi _{\Delta }^{*(1)}-\frac{\delta
^{L}\bar{\psi}_{1}}{\delta \Phi ^{\Delta }}\right) \mu _{1}^{\Delta }+\left( 
\bar{\Phi}_{\Delta }^{(2)}-\frac{\delta ^{L}\bar{\psi}_{1}}{\delta \mu
_{3}^{\Delta }}\right) \nu _{2}^{\Delta }+  \nonumber \\
&&\left( \bar{\Phi}_{\Delta }^{(3)}+\frac{\delta ^{L}\bar{\psi}_{1}}{\delta
\mu _{2}^{\Delta }}\right) \nu _{3}^{\Delta }+\left( \tilde{\Phi}_{\Delta }-%
\frac{\delta ^{L}\bar{\psi}_{1}}{\delta \nu _{1}^{\Delta }}\right) \omega
^{\Delta },  \label{sp3.142}
\end{eqnarray}
where the Lagrange multipliers have the properties 
\begin{equation}
\mu _{1}^{\Delta }=\left( \mu _{1}^{A},m_{1}^{I}\right) ,\;\nu _{2}^{\Delta
}=\left( \nu _{2}^{A},n_{2}^{I}\right) ,\;\nu _{3}^{\Delta }=\left( \nu
_{3}^{A},n_{3}^{I}\right) ,\;\omega ^{\Delta }=\left( \omega
^{A},o^{I}\right) ,  \label{sp3.143}
\end{equation}
\begin{equation}
\epsilon \left( \mu _{1}^{\Delta }\right) =\epsilon \left( \omega ^{\Delta
}\right) =\epsilon \left( \Phi ^{\Delta }\right) +1,\;\epsilon \left( \nu
_{2}^{\Delta }\right) =\epsilon \left( \nu _{3}^{\Delta }\right) =\epsilon
\left( \Phi ^{\Delta }\right) .  \label{sp3.144}
\end{equation}
Taking into account the formulas (\ref{sp3.135}), (\ref{sp3.112}--\ref
{sp3.115}) with $A$, $B$, $C$ replaced by $\Delta $, $\Delta ^{\prime }$, $%
\Delta ^{\prime \prime }$, and $\psi $ by $\bar{\psi}^{\prime }$, we find
that the effective action reads as 
\begin{eqnarray}
&&S_{\mathrm{eff}}=\bar{S}+\sum\limits_{a=1}^{3}\left( \Phi _{\Delta
}^{*(a)}\mu _{a}^{\Delta }+\bar{\Phi}_{\Delta }^{(a)}\nu _{a}^{\Delta
}\right) +\left( \tilde{\Phi}_{\Delta }-\frac{\delta ^{L}\bar{\psi}^{\prime }%
}{\delta \Phi ^{\Delta }}\right) \omega ^{\Delta }+  \nonumber \\
&&\left( -\right) ^{\epsilon \left( \Phi ^{\Delta ^{\prime }}\right) }\frac{%
\delta ^{3L}\bar{\psi}^{\prime }}{\delta \Phi ^{\Delta }\delta \Phi ^{\Delta
^{\prime }}\delta \Phi ^{\Delta ^{\prime \prime }}}\mu _{3}^{\Delta ^{\prime
\prime }}\mu _{2}^{\Delta ^{\prime }}\mu _{1}^{\Delta }-\frac{\delta ^{2L}%
\bar{\psi}^{\prime }}{\delta \Phi ^{\Delta }\delta \Phi ^{\Delta ^{\prime }}}%
\sum\limits_{a=1}^{3}\left( \nu _{a}^{\Delta ^{\prime }}\mu _{a}^{\Delta
}\right) ,  \label{sp3.145}
\end{eqnarray}
so the corresponding path integral is of the type 
\begin{equation}
Z_{\mathrm{eff}}=\int \mathcal{D}\Phi ^{\Delta }\mathcal{D}\Phi _{\Delta
}^{*(a)}\mathcal{D}\bar{\Phi}_{\Delta }^{(a)}\mathcal{D}\tilde{\Phi}_{\Delta
}\mathcal{D}\mu _{a}^{\Delta }\mathcal{D}\nu _{a}^{\Delta }\mathcal{D}\omega
^{\Delta }\exp \left( iS_{\mathrm{eff}}\right) .  \label{sp3.146}
\end{equation}
If one integrates in $Z_{\mathrm{eff}}$ over the auxiliary variables $\mu
_{1}^{\Delta }$, $\nu _{2}^{\Delta }$, $\nu _{3}^{\Delta }$, $\omega
^{\Delta }$, $\Phi _{\Delta }^{*(1)}$, $\bar{\Phi}_{\Delta }^{(2)}$, $\bar{%
\Phi}_{\Delta }^{(3)}$ and $\tilde{\Phi}_{\Delta }$, one reobtains the path
integral (\ref{sp3.141}).

Let us work in the sequel with the second antibracket and forget about the
other two. In this case we need to introduce the variables $\left( \mu
_{1}^{\Delta },\mu _{3}^{\Delta },\nu _{2}^{\Delta }\right) $ respectively
conjugated in the second antibracket to $\left( \bar{\Phi}_{\Delta }^{(3)},%
\bar{\Phi}_{\Delta }^{(1)},\tilde{\Phi}_{\Delta }\right) $, and work with
the gauge-fixing fermion 
\begin{equation}
\bar{\psi}_{2}=\left( -\right) ^{1+\epsilon \left( \Phi ^{\Lambda }\right)
\left( \epsilon \left( \Phi ^{\Delta }\right) +1\right) }\frac{\delta ^{2L}%
\bar{\psi}^{\prime }\left[ \Phi ^{\Delta }\right] }{\delta \Phi ^{\Delta
}\delta \Phi ^{\Lambda }}\mu _{1}^{\Delta }\mu _{3}^{\Lambda }+\frac{\delta
^{L}\bar{\psi}^{\prime }\left[ \Phi ^{\Delta }\right] }{\delta \Phi ^{\Delta
}}\nu _{2}^{\Delta },  \label{sp3.147}
\end{equation}
while the gauge-fixing conditions resulting from $\bar{\psi}_{2}$ are
obtained with the help of the same $\bar{\psi}^{\prime }$ like in (\ref
{sp3.136}), and will be enforced within the effective action through the
Lagrange multipliers $\left( \mu _{2}^{\Delta },\nu _{1}^{\Delta },\nu
_{3}^{\Delta },\omega ^{\Delta }\right) $. Similarly, if we focus on the
third antibracket, then we add $\left( \mu _{1}^{\Delta },\mu _{2}^{\Delta
},\nu _{3}^{\Delta }\right) $, which are taken to be conjugated in the third
antibracket to $\left( \bar{\Phi}_{\Delta }^{(2)},\bar{\Phi}_{\Delta }^{(1)},%
\tilde{\Phi}_{\Delta }\right) $, and consider the gauge-fixing fermion 
\begin{equation}
\bar{\psi}_{3}=\left( -\right) ^{1+\epsilon \left( \Phi ^{\Lambda }\right)
\left( \epsilon \left( \Phi ^{\Delta }\right) +1\right) }\frac{\delta ^{2L}%
\bar{\psi}^{\prime }\left[ \Phi ^{\Delta }\right] }{\delta \Phi ^{\Delta
}\delta \Phi ^{\Lambda }}\mu _{2}^{\Delta }\mu _{1}^{\Lambda }+\frac{\delta
^{L}\bar{\psi}^{\prime }\left[ \Phi ^{\Delta }\right] }{\delta \Phi ^{\Delta
}}\nu _{3}^{\Delta }.  \label{sp3.148}
\end{equation}
The consequent gauge-fixing conditions are implemented within the effective
action via the Lagrange multipliers $\left( \mu _{3}^{\Delta },\nu
_{1}^{\Delta },\nu _{2}^{\Delta },\omega ^{\Delta }\right) $. Thus, we can
derive the path integral (\ref{sp3.146}) working with any of the
antibrackets, so we conclude that (\ref{sp3.146}) is $s_{a}$-invariant. This
follows from the general properties of the formalism developed until now.
One can also check by direct computation that the effective action (\ref
{sp3.145}) is invariant under the `gauge-fixed' BRST $Sp(3)$ transformations 
\begin{equation}
s_{a}\Phi ^{\Delta }=\left( -\right) ^{\epsilon \left( \Phi ^{\Delta
}\right) }\mu _{a}^{\Delta },  \label{sp3.151}
\end{equation}
\begin{equation}
s_{a}\Phi _{\Delta }^{*(b)}=\delta _{ab}\frac{\delta ^{R}\bar{S}}{\delta
\Phi ^{\Delta }},  \label{sp3.152}
\end{equation}
\begin{equation}
s_{a}\bar{\Phi}_{\Delta }^{(b)}=\varepsilon _{abc}\Phi _{\Delta }^{*(c)},
\label{sp3.153}
\end{equation}
\begin{equation}
s_{a}\tilde{\Phi}_{\Delta }=\bar{\Phi}_{\Delta }^{(a)},  \label{sp3.154}
\end{equation}
\begin{equation}
s_{a}\mu _{b}^{\Delta }=\left( -\right) ^{\epsilon \left( \Phi ^{\Delta
}\right) }\varepsilon _{abc}\nu _{c}^{\Delta },  \label{sp3.155}
\end{equation}
\begin{equation}
s_{a}\nu _{b}^{\Delta }=\left( -\right) ^{\epsilon \left( \Phi ^{\Delta
}\right) }\delta _{ab}\omega ^{\Delta },  \label{sp3.156}
\end{equation}
\begin{equation}
s_{a}\omega ^{\Delta }=0,  \label{sp3.157}
\end{equation}
up to some `skew-symmetric' combinations of equations of motion. In this
way, the announced aim, of obtaining an effective action that is $s_{a}$%
-invariant, has been accomplished. This completes our formalism.

\section{Example}

Let us exemplify the general theory developed so far. In order to emphasize
the key features of our method, we consider the simple case of abelian gauge
fields. We begin with the Lagrangian action 
\begin{equation}
S_{0}\left[ A^{\alpha }\right] =-\frac{1}{4}\int d^{4}xF_{\alpha \beta
}F^{\alpha \beta },  \label{sp3.e1}
\end{equation}
invariant under the abelian and irreducible gauge transformations $\delta
_{\varepsilon }A^{\alpha }=\partial ^{\alpha }\varepsilon $, where $%
F_{\alpha \beta }=\partial _{\alpha }A_{\beta }-\partial _{\beta }A_{\alpha }
$. The analogies between the general formalism and this example are: $\alpha
_{0}\rightarrow \alpha $, $\Phi ^{\alpha _{0}}\rightarrow A^{\alpha }$, $%
\alpha _{1}\rightarrow 1$, $\varepsilon ^{\alpha _{1}}\rightarrow
\varepsilon $, $Z_{\;\;\alpha _{1}}^{\alpha _{0}}\rightarrow \partial
^{\alpha }$. According to the general ideas exposed in Section 3, we
triplicate the gauge parameters, and consider the second-stage reducible
gauge transformations $\delta _{\varepsilon }A^{\alpha }=\partial ^{\alpha
}\varepsilon _{1}+\partial ^{\alpha }\varepsilon _{2}+\partial ^{\alpha
}\varepsilon _{3}$. Then, the ghost spectrum reads as 
\begin{equation}
\stackrel{(1,0,0)}{\eta }_{1},\stackrel{(0,1,0)}{\eta }_{2},\stackrel{(0,0,1)%
}{\eta }_{3},\stackrel{(0,1,1)}{\pi }_{1},\stackrel{(1,0,1)}{\pi }_{2},%
\stackrel{(1,1,0)}{\pi }_{3},\stackrel{(1,1,1)}{\lambda },  \label{sp3.e4}
\end{equation}
where 
\begin{equation}
\epsilon \left( \eta _{a}\right) =\epsilon \left( \lambda \right)
=1,\;\epsilon \left( \pi _{a}\right) =0,\;a=1,2,3.  \label{sp3.e5}
\end{equation}
The antifield spectrum and its properties result from the general line
exposed in sections 4 and 5. The solution to the classical master equation (%
\ref{sp3.68}) of the $Sp(3)$ formalism reads as 
\begin{eqnarray}
&&S=\int d^{4}x\left( -\frac{1}{4}F_{\alpha \beta }F^{\alpha \beta
}+A_{\alpha }^{*(a)}\partial ^{\alpha }\eta _{a}+\varepsilon _{abc}\eta
_{b}^{*(a)}\pi _{c}+\right.   \nonumber \\
&&\left. \bar{A}_{\alpha }^{(c)}\partial ^{\alpha }\pi _{c}-\left( \pi
_{a}^{*(a)}+\bar{\eta}_{a}^{(a)}\right) \lambda +\tilde{A}_{\alpha }\partial
^{\alpha }\lambda \right) ,  \label{sp3.e19}
\end{eqnarray}
and it actually reduces to the boundary conditions (\ref{sp3.71}--\ref
{sp3.73}).

Initially, we will fix the gauge without introducing the supplementary
fields $\varphi ^{\alpha _{1}}$. In view of this, we pass to the first
antibracket and discard the other two. Consequently, we need the additional
variables (\ref{sp3.100a}) in order to obtain that every variable has its
conjugated in the first antibracket. Thus, the solution $S_{1}$ to the
standard master equation in the first antibracket (\ref{sp3.109b}) reads as 
\begin{eqnarray}
&&S_{1}=S+\int d^{4}x\left( A_{\alpha }^{*(2)}\mu _{2}^{(A)\alpha
}+\sum\limits_{a=1}^{3}\left( \eta _{a}^{*(2)}\mu _{2}^{(\eta _{a})}+\pi
_{a}^{*(2)}\mu _{2}^{(\pi _{a})}\right) +\lambda ^{*(2)}\mu _{2}^{(\lambda
)}+\right.  \nonumber \\
&&A_{\alpha }^{*(3)}\mu _{3}^{(A)\alpha }+\sum\limits_{a=1}^{3}\left( \eta
_{a}^{*(3)}\mu _{3}^{(\eta _{a})}+\pi _{a}^{*(3)}\mu _{3}^{(\pi
_{a})}\right) +\lambda ^{*(3)}\mu _{3}^{(\lambda )}+\bar{A}_{\alpha
}^{(1)}\nu _{1}^{(A)\alpha }+  \nonumber \\
&&\left. \sum\limits_{a=1}^{3}\left( \bar{\eta}_{a}^{(1)}\nu _{1}^{(\eta
_{a})}+\bar{\pi}_{a}^{(1)}\nu _{1}^{(\pi _{a})}\right) +\bar{\lambda}%
^{(1)}\nu _{1}^{(\lambda )}\right) ,  \label{sp3.e26}
\end{eqnarray}
where we put an extra upper index for distinguishing between the various
types of $\mu $'s and $\nu $'s associated with different fields that carry
the same index $A$. We employ a gauge-fixing fermion of the type (\ref
{sp3.111}) for performing the gauge-fixing procedure. As we need to choose a
definite form for the fermionic functional $\psi $, let us examine a little
bit the form of $S_{1}$. The `gauge-fixing' term for the electromagnetic
field is yielded via the elimination of $\tilde{A}_{\alpha }$ from the term $%
\tilde{A}_{\alpha }\partial ^{\alpha }\lambda $ present into the solution $S$%
. Then, we have to put in $\psi $ a term that is at least quadratic in $%
A^{\alpha }$, such that $\tilde{A}_{\alpha }=\delta ^{L}\psi /\delta
A^{\alpha }$ effectively depends on Maxwell's field. For definiteness, we
choose this term to be precisely quadratically, $\psi \sim A^{\alpha
}A_{\alpha }$. As $\psi $ is fermionic, the term $A^{\alpha }A_{\alpha }$
should be multiplied by a scalar fermionic ghost. There are actually four
ghosts that satisfy this requirement, namely, $\left( \eta _{a}\right)
_{a=1,2,3}$ and $\lambda $. As the ghost of ghost of ghost $\lambda $ is
already implied in $\tilde{A}_{\alpha }\partial ^{\alpha }\lambda $, we are
tempted to take $\psi \sim A^{\alpha }A_{\alpha }\lambda $ for forcing the
couplings in the gauge-fixed action to involve less fields/ghosts. In the
meantime, we avoid other terms in $\psi $ for not complicating unnecessarily
the gauge-fixed action. In conclusion, we try $\psi $ under the form 
\begin{equation}
\psi =\int d^{4}x\left( \frac{1}{2}A^{\alpha }A_{\alpha }\lambda \right) ,
\label{sp3.e27}
\end{equation}
with the help of which we infer (see formulas (\ref{sp3.112}--\ref{sp3.115}%
)) 
\begin{equation}
A_{\alpha }^{*(1)}=\mu _{3\alpha }^{(A)}\mu _{2}^{(\lambda )}-\mu _{2\alpha
}^{(A)}\mu _{3}^{(\lambda )}+\lambda \nu _{1}^{(A)\alpha }+A_{\alpha }\nu
_{1}^{(\lambda )},  \label{sp3.e28}
\end{equation}
\begin{equation}
\eta _{1}^{*(1)}=\eta _{2}^{*(1)}=\eta _{3}^{*(1)}=\pi _{1}^{*(1)}=\pi
_{2}^{*(1)}=\pi _{3}^{*(1)}=0,  \label{sp3.e29}
\end{equation}
\begin{equation}
\lambda ^{*(1)}=-\mu _{3\alpha }^{(A)}\mu _{2}^{(A)\alpha }+A_{\alpha }\nu
_{1}^{(A)\alpha },\;\bar{A}_{\alpha }^{(2)}=\lambda \mu _{2\alpha
}^{(A)}+A_{\alpha }\mu _{2}^{(\lambda )},  \label{sp3.e30}
\end{equation}
\begin{equation}
\bar{\eta}_{1}^{(2)}=\bar{\eta}_{2}^{(2)}=\bar{\eta}_{3}^{(2)}=\bar{\pi}%
_{1}^{(2)}=\bar{\pi}_{2}^{(2)}=\bar{\pi}_{3}^{(2)}=0,  \label{sp3.e31}
\end{equation}
\begin{equation}
\bar{\lambda}^{(2)}=-A_{\alpha }\mu _{2}^{(A)\alpha },\;\bar{A}_{\alpha
}^{(3)}=\lambda \mu _{3\alpha }^{(A)}+A_{\alpha }\mu _{3}^{(\lambda )},
\label{sp3.e32}
\end{equation}
\begin{equation}
\bar{\eta}_{1}^{(3)}=\bar{\eta}_{2}^{(3)}=\bar{\eta}_{3}^{(3)}=\bar{\pi}%
_{1}^{(3)}=\bar{\pi}_{2}^{(3)}=\bar{\pi}_{3}^{(3)}=0,  \label{sp3.e33}
\end{equation}
\begin{equation}
\bar{\lambda}^{(3)}=-A_{\alpha }\mu _{3}^{(A)\alpha },\;\tilde{A}_{\alpha
}=A_{\alpha }\lambda ,  \label{sp3.e34}
\end{equation}
\begin{equation}
\tilde{\eta}_{1}=\tilde{\eta}_{2}=\tilde{\eta}_{3}=\tilde{\pi}_{1}=\tilde{\pi%
}_{2}=\tilde{\pi}_{3}=0,\;\tilde{\lambda}=\frac{1}{2}A^{\alpha }A_{\alpha }.
\label{sp3.e35}
\end{equation}
Substituting the above relations in (\ref{sp3.e26}), we obtain a gauge-fixed
action of the type (\ref{sp3.116}). Integrating in the resulting gauge-fixed
path integral over some auxiliary fields, we arrive at 
\begin{equation}
Z_{1\psi }=\int \mathcal{D}A^{\alpha }\mathcal{D}\eta _{a}\mathcal{D}\pi _{a}%
\mathcal{D}\lambda \exp \left( iS_{1\psi }^{\prime }\right) ,
\label{sp3.e36}
\end{equation}
where 
\begin{equation}
S_{1\psi }^{\prime }=\int d^{4}x\left( -\frac{1}{4}F_{\alpha \beta
}F^{\alpha \beta }-\sum\limits_{a=1}^{3}\lambda \left( \partial _{\alpha
}\pi _{a}\right) \left( \partial ^{\alpha }\eta _{a}\right) +\lambda
A_{\alpha }\partial ^{\alpha }\lambda \right) .  \label{sp3.e37}
\end{equation}
All the terms in the gauge-fixed action excluding those appearing in the
original Lagrangian action describe couplings of order three among fields
and ghosts. Now, it appears clearly that we cannot reach the familiar form
of the gauge-fixed action for Maxwell's theory involving a term $b\partial
_{\alpha }A^{\alpha }$ that enforces the Lorentz gauge condition $\partial
_{\alpha }A^{\alpha }=0$, or a Gaussian term of the type $b\left( \partial
_{\alpha }A^{\alpha }+\frac{1}{2}b\right) $, where $b$ is an auxiliary
bosonic field.

In the sequel, we will fix the gauge following the more elaborated line
exposed in the final part of Section 7. We intend to implement the
gauge-fixing conditions $f_{\alpha _{1}}\left( \Phi ^{\alpha _{0}}\right)
=0\rightarrow \partial _{\alpha }A^{\alpha }=0$. In view of this, we add a
fermionic scalar field $\varphi $ subject to the gauge transformation $%
\delta _{\xi }\varphi =\xi $. The solution to the classical master equation
of the $Sp(3)$ BRST formalism for the overall gauge theory has the form 
\begin{equation}
\bar{S}=S+\int d^{4}x\left( \varphi ^{*(a)}C_{a}+\left( \varepsilon
_{abc}C_{b}^{*(a)}+\bar{\varphi}^{(c)}\right) p_{c}-\left( p_{a}^{*(a)}+\bar{%
C}_{a}^{(a)}-\tilde{\varphi}\right) l\right) ,  \label{sp3.e52a}
\end{equation}
(see (\ref{sp3.132}) where the index $\alpha _{1}$ has been suppressed). In
order to pass to the solution of the master equation in the first
antibracket, we enlarge the field and antifield spectra with the fields of
the type $r$, $k$, $m$ and $n$, such that we arrive at (see (\ref{sp3.133})) 
\begin{eqnarray}
&&\bar{S}_{1}=\bar{S}+\int d^{4}x\left( A_{\alpha }^{*(2)}\mu
_{2}^{(A)\alpha }+\sum\limits_{a=1}^{3}\left( \eta _{a}^{*(2)}\mu
_{2}^{(\eta _{a})}+\pi _{a}^{*(2)}\mu _{2}^{(\pi _{a})}\right) +\lambda
^{*(2)}\mu _{2}^{(\lambda )}+\right.  \nonumber \\
&&A_{\alpha }^{*(3)}\mu _{3}^{(A)\alpha }+\sum\limits_{a=1}^{3}\left( \eta
_{a}^{*(3)}\mu _{3}^{(\eta _{a})}+\pi _{a}^{*(3)}\mu _{3}^{(\pi
_{a})}\right) +\lambda ^{*(3)}\mu _{3}^{(\lambda )}+\bar{A}_{\alpha
}^{(1)}\nu _{1}^{(A)\alpha }+  \nonumber \\
&&\sum\limits_{a=1}^{3}\left( \bar{\eta}_{a}^{(1)}\nu _{1}^{(\eta _{a})}+%
\bar{\pi}_{a}^{(1)}\nu _{1}^{(\pi _{a})}\right) +\bar{\lambda}^{(1)}\nu
_{1}^{(\lambda )}+\varphi ^{*(2)}m_{2}^{(\varphi
)}+l^{*(2)}m_{2}^{(l)}+\varphi ^{*(3)}m_{3}^{(\varphi )}+  \nonumber \\
&&\sum\limits_{a=1}^{3}\left(
C_{a}^{*(2)}m_{2}^{(C_{a})}+p_{a}^{*(2)}m_{2}^{(p_{a})}\right)
+\sum\limits_{a=1}^{3}\left(
C_{a}^{*(3)}m_{3}^{(C_{a})}+p_{a}^{*(3)}m_{3}^{(p_{a})}\right) +  \nonumber
\\
&&\left. l^{*(3)}m_{3}^{(l)}+\bar{\varphi}^{(1)}n_{1}^{(\varphi
)}+\sum\limits_{a=1}^{3}\left( \bar{C}_{a}^{(1)}n_{1}^{(C_{a})}+\bar{p}%
_{a}^{(1)}n_{1}^{(p_{a})}\right) +\bar{l}^{(1)}n_{1}^{(l)}\right) .
\label{sp3.e59}
\end{eqnarray}
The gauge-fixing process relies on the choice of a certain fermionic
functional of the type (\ref{sp3.135}), with the fermion $\bar{\psi}^{\prime
}$ like in (\ref{sp3.136}). We take $\bar{\psi}\left[ \Phi ^{\Delta }\right]
\rightarrow \int d^{4}x\left( A^{\alpha }\partial _{\alpha }\lambda +\frac{1%
}{2}\varphi l\right) $ and $f_{\alpha _{1}}\left( \Phi ^{\alpha _{0}}\right)
\rightarrow \partial _{\alpha }A^{\alpha }$, such that 
\begin{equation}
\bar{\psi}^{\prime }=\int d^{4}x\left( A^{\alpha }\partial _{\alpha }\lambda
+\varphi \left( \frac{1}{2}l+\partial _{\alpha }A^{\alpha }\right) \right) .
\label{sp3.e60}
\end{equation}
Eliminating some `fields' and `antifields' from (\ref{sp3.e59}) with the
help of (\ref{sp3.e60}), as explained in Section 7, and further integrating
in the resulting path integral over some auxiliary variables, we finally
deduce 
\begin{equation}
Z_{1\bar{\psi}^{\prime }}=\int \mathcal{D}A^{\alpha }\mathcal{D}\varphi 
\mathcal{D}\eta _{a}\mathcal{D}C_{a}\mathcal{D}\pi _{a}\mathcal{D}p_{a}%
\mathcal{D}\lambda \mathcal{D}l\exp \left( i\bar{S}_{1\bar{\psi}^{\prime
}}^{\prime }\right) ,  \label{sp3.e61}
\end{equation}
where 
\begin{eqnarray}
&&\bar{S}_{1\bar{\psi}^{\prime }}^{\prime }=\int d^{4}x\left( -\frac{1}{4}%
F_{\alpha \beta }F^{\alpha \beta }+\left( \frac{1}{2}l+\partial _{\alpha
}A^{\alpha }\right) l-\left( \partial _{\alpha }\varphi \right) \left(
\partial ^{\alpha }\lambda \right) +\right.  \nonumber \\
&&\left. \sum\limits_{a=1}^{3}\left( \left( \partial _{\alpha }p_{a}\right)
\left( \partial ^{\alpha }\eta _{a}\right) +\left( \partial ^{\alpha
}C_{a}\right) \left( \partial _{\alpha }\pi _{a}\right) \right) \right) .
\label{sp3.e62}
\end{eqnarray}
We observe that we were able to enforce a Gaussian term in the gauge-fixed
action, of the type $\left( \frac{1}{2}l+\partial _{\alpha }A^{\alpha
}\right) l$, as the ghost of ghost of ghost $l$ corresponding to the
supplementary scalar fermionic field $\varphi $ is bosonic. If we want to
implement the Lorentz gauge-fixing condition $\partial _{\alpha }A^{\alpha
}=0$, it is enough to discard the term $\frac{1}{2}\varphi l$ from (\ref
{sp3.e60}), obtaining thus $\bar{S}_{1\bar{\psi}^{\prime }}^{\prime }$
without the term $\frac{1}{2}\left( l\right) ^{2}$. In the meantime, we
notice that $\bar{\psi}^{\prime }$ is quadratic in the fields, which was not
possible for a fermion of the type $\psi $ (which must be at least cubic,
see (\ref{sp3.e27})). Finally, we remark that the gauge-fixed action (\ref
{sp3.e62}) bear the trace of the triplication. Indeed, the terms $\left(
\partial _{\alpha }p_{a}\right) \left( \partial ^{\alpha }\eta _{a}\right) $
correspond to the triplication of the gauge transformations, $\left(
\partial _{\alpha }\pi _{a}\right) \left( \partial ^{\alpha }C_{a}\right) $
are associated with the induced first-stage reducibility relations, while $%
\left( \partial _{\alpha }\varphi \right) \left( \partial ^{\alpha }\lambda
\right) $ are correlated with the second-stage reducibility relations. In
this context, $p_{a}$, $C_{a}$ and $\varphi $ play the role of Lagrangian
antighosts respectively coresponding to $\eta _{a}$, $\pi _{a}$ and $\lambda 
$. This completes the analysis of the model under study.

\section{Conclusion}

To conclude with, in this paper we have shown that the $Sp(3)$ BRST symmetry
for irreducible theories can be developed in the antibracket-antifield
formulation by adapting the methods of homological perturbation theory. The
key point of our approach is the construction of a Koszul-Tate triresolution
of the algebra of smooth functions defined on the stationary surface of
field equations, that allows us to apply a positivity-like theorem for
triresolutions. With the canonical generator of the $Sp(3)$ BRST symmetry at
hand, we give a gauge-fixing procedure specific to the standard
antibracket-antifield formalism, that leads to an effective action which is $%
s_{a}$-invariant. The general procedure is finally applied on abelian gauge
fields.

\section*{Acknowledgment}

This work has been supported by a Romanian National Council for Academic
Scientific Research (CNCSIS) grant.

\appendix 

\section{Appendix: basic properties of triresolutions}

\begin{definition}
Let $\mathcal{A}_{0}$ be an algebra and $\mathcal{A}^{\prime }$ be a
trigraded algebra with the tridegree called resolution tridegree, $\mathrm{%
trires}=\left( \mathrm{res}_{1},\mathrm{res}_{2},\mathrm{res}_{3}\right) $,
where all $\mathrm{res}_{a}$ ($a=1,2,3$) are assumed to be nonnegative
integers, $\mathrm{res}_{a}\geq 0$, $a=1,2,3$. We define the total
resolution degree as $\mathrm{res}=\mathrm{res}_{1}+\mathrm{res}_{2}+\mathrm{%
res}_{3}$. Let $\delta :\mathcal{A}^{\prime }\rightarrow \mathcal{A}^{\prime
}$ be a differential of total resolution degree minus one, $\delta ^{2}=0$, $%
\mathrm{res}\left( \delta \right) =-1$, such that $\mathrm{res}\left( \delta
a\right) =\mathrm{res}\left( a\right) -1$ when $\mathrm{res}\left( a\right)
\geq 1$, or $\mathrm{res}\left( \delta a\right) =0$ when $\mathrm{res}\left(
a\right) =0$, in which case $\delta a=0$. One says that the simple
differential complex $\left( \delta ,\mathcal{A}^{\prime }\right) $ induces
a triresolution $\left( \left( \delta _{a}\right) _{a=1,2,3},\mathcal{A}%
^{\prime }\right) $ of the algebra $\mathcal{A}_{0}$ if and only if:

1. The differential $\delta $ splits as the sum among three derivations
only, $\delta =\delta _{1}+\delta _{2}+\delta _{3}$, with $\mathrm{trires}%
\left( \delta _{1}\right) =\left( -1,0,0\right) $, $\mathrm{trires}\left(
\delta _{2}\right) =\left( 0,-1,0\right) $, and $\mathrm{trires}\left(
\delta _{3}\right) =\left( 0,0,-1\right) $ (no extra piece, say, of
resolution tridegree $\left( -2,1,0\right) $ occurs). The nilpotency of $%
\delta $ induces that $\left( \delta _{a}\right) _{a=1,2,3}$ are three
anticommuting differentials, $\delta _{a}\delta _{b}+\delta _{b}\delta
_{a}=0 $, $a,b=1,2,3$.

2. We have that 
\begin{equation}
H_{0,0,0}\left( \delta _{a}\right) =\mathcal{A}_{0},\;H_{i,j,k}\left( \delta
_{a}\right) =0,\;a=1,2,3,\;i,j,k\geq 0,\;i+j+k>0,  \label{sp3.ap1}
\end{equation}
\begin{equation}
H_{0}\left( \delta \right) =\mathcal{A}_{0},\;H_{l}\left( \delta \right)
=0,\;l>0,  \label{sp3.ap2}
\end{equation}
where $H_{i,j,k}\left( \delta _{a}\right) $ signifies the space of elements
with $\mathrm{trires}=(i,j,k)$, that are $\delta _{a}$-closed modulo $\delta
_{a}$-exact, and $H_{l}\left( \delta \right) $ means the cohomological space
spanned by the objects with $\mathrm{res}=l$, that are $\delta $-closed
modulo $\delta $-exact.
\end{definition}

\begin{theorem}
Let $\left( \left( \delta _{a}\right) _{a=1,2,3},\mathcal{A}^{\prime
}\right) $ be a triresolution and $\stackrel{[i,j,k]}{F}\in \mathcal{A}%
^{\prime }$ (assuming $i+j+k>0$), be such that 
\begin{equation}
\left( \delta _{a}\stackrel{[i,j,k]}{F}=0,\;a=1,2,3\right) \Leftrightarrow
\delta \stackrel{[i,j,k]}{F}=0.  \label{sp3.ap3}
\end{equation}
Then, $\stackrel{[i,j,k]}{F}$ can be represented as 
\begin{equation}
\stackrel{\lbrack i,j,k]}{F}=\delta _{3}\delta _{2}\delta _{1}\stackrel{%
[i+1,j+1,k+1]}{F}.  \label{sp3.ap4}
\end{equation}
\end{theorem}

{\textbf{Proof.}} From (\ref{sp3.ap3}) it follows that $\delta _{3}\stackrel{%
[i,j,k]}{F}=0$, which further yields that there exists an element $\stackrel{%
[i,j,k+1]}{R}$ such that 
\begin{equation}
\stackrel{\lbrack i,j,k]}{F}=\delta _{3}\stackrel{[i,j,k+1]}{R},
\label{sp3.ap5}
\end{equation}
since $H_{i,j,k}\left( \delta _{3}\right) =0$ for $i+j+k>0$. But one also
has $\delta _{2}\stackrel{[i,j,k]}{F}=0$, hence $\delta _{2}\delta _{3}%
\stackrel{[i,j,k+1]}{R}=0$, which is equivalent to $\delta _{3}\left( \delta
_{2}\stackrel{[i,j,k+1]}{R}\right) =0$ on account of the anticommutativity
between $\delta _{2}$ and $\delta _{3}$. This means that there exists an
object $\stackrel{[i,j-1,k+2]}{R^{\prime }}$ such that 
\begin{equation}
\delta _{2}\stackrel{[i,j,k+1]}{R}=\delta _{3}\stackrel{[i,j-1,k+2]}{%
R^{\prime }}.  \label{sp3.ap6}
\end{equation}
Similarly, we have that $\delta _{1}\stackrel{[i,j,k]}{F}=0$, which induces
that $\delta _{3}\left( \delta _{1}\stackrel{[i,j,k+1]}{R}\right) =0$ by
means of the anticommutativity between $\delta _{1}$ and $\delta _{3}$.
Therefore, there exists a certain $\stackrel{[i-1,j,k+2]}{R^{\prime \prime }}
$ with the property 
\begin{equation}
\delta _{1}\stackrel{[i,j,k+1]}{R}=\delta _{3}\stackrel{[i-1,j,k+2]}{%
R^{\prime \prime }}.  \label{sp3.ap7}
\end{equation}
If we apply: (i) $\delta _{2}$ on (\ref{sp3.ap6}); (ii) $\delta _{1}$ on (%
\ref{sp3.ap7}); (iii) $\delta _{1}$ on (\ref{sp3.ap6}), $\delta _{2}$ on (%
\ref{sp3.ap7}), and add the results, then we find the equations 
\begin{equation}
\delta _{2}\stackrel{[i,j-1,k+2]}{R^{\prime }}=\delta _{3}\stackrel{%
[i,j-2,k+3]}{R^{\prime }},  \label{sp3.ap8}
\end{equation}
\begin{equation}
\delta _{1}\stackrel{[i-1,j,k+2]}{R^{\prime \prime }}=\delta _{3}\stackrel{%
[i-2,j,k+3]}{R^{\prime \prime }},  \label{sp3.ap9}
\end{equation}
\begin{equation}
\delta _{1}\stackrel{[i,j-1,k+2]}{R^{\prime }}+\delta _{2}\stackrel{%
[i-1,j,k+2]}{R^{\prime \prime }}=\delta _{3}\stackrel{[i-1,j-1,k+3]}{Q}.
\label{sp3.ap10}
\end{equation}
Next, we perform the following steps: we apply (i) $\delta _{2}$ on (\ref
{sp3.ap8}); (ii) $\delta _{1}$ on (\ref{sp3.ap9}); (iii) $\delta _{1}$ on (%
\ref{sp3.ap8}), $\delta _{2}$ on (\ref{sp3.ap10}), and add the resulting
equations; (iv) $\delta _{2}$ on (\ref{sp3.ap9}), $\delta _{1}$ on (\ref
{sp3.ap10}), and add the resulting equations, then we get 
\begin{equation}
\delta _{2}\stackrel{[i,j-2,k+3]}{R^{\prime }}=\delta _{3}\stackrel{%
[i,j-3,k+4]}{R^{\prime }},  \label{sp3.ap11}
\end{equation}
\begin{equation}
\delta _{1}\stackrel{[i-2,j,k+3]}{R^{\prime \prime }}=\delta _{3}\stackrel{%
[i-3,j,k+4]}{R^{\prime \prime }},  \label{sp3.ap12}
\end{equation}
\begin{equation}
\delta _{1}\stackrel{[i,j-2,k+3]}{R^{\prime }}+\delta _{2}\stackrel{%
[i-1,j-1,k+3]}{Q}=\delta _{3}\stackrel{[i-1,j-2,k+4]}{Q^{\prime }}
\label{sp3.ap13}
\end{equation}
\begin{equation}
\delta _{2}\stackrel{[i-2,j,k+3]}{R^{\prime \prime }}+\delta _{1}\stackrel{%
[i-1,j-1,k+3]}{Q}=\delta _{3}\stackrel{[i-2,j-1,k+4]}{Q^{\prime \prime }}.
\label{sp3.ap14}
\end{equation}
From now on, we obviously derive a tower of descent equations of the form 
\begin{equation}
\delta _{2}\stackrel{[i,j-m,k+m+1]}{R^{\prime }}=\delta _{3}\stackrel{%
[i,j-m-1,k+m+2]}{R^{\prime }},  \label{sp3.ap15}
\end{equation}
\begin{equation}
\delta _{1}\stackrel{[i-m,j,k+m+1]}{R^{\prime \prime }}=\delta _{3}\stackrel{%
[i-m-1,j,k+m+2]}{R^{\prime \prime }},  \label{sp3.ap16}
\end{equation}
\begin{equation}
\delta _{1}\stackrel{[i,j-m,k+m+1]}{R^{\prime }}+\delta _{2}\stackrel{%
[i-1,j-m+1,k+m+1]}{Q^{\prime }}=\delta _{3}\stackrel{[i-1,j-m,k+m+2]}{%
Q^{\prime }},  \label{sp3.ap17}
\end{equation}
\begin{equation}
\delta _{2}\stackrel{[i-m,j,k+m+1]}{R^{\prime \prime }}+\delta _{1}\stackrel{%
[i-m+1,j-1,k+m+1]}{Q^{\prime \prime }}=\delta _{3}\stackrel{[i-m,j-1,k+m+2]}{%
Q^{\prime \prime }},  \label{sp3.ap18}
\end{equation}
for $m\geq 3$. The last equations in (\ref{sp3.ap17}--\ref{sp3.ap18}) read
as 
\begin{equation}
\delta _{2}\stackrel{[i-1,0,k+j+2]}{Q^{\prime }}=0,\;m=j+1,  \label{sp3.ap19}
\end{equation}
\begin{equation}
\delta _{1}\stackrel{[0,j-1,k+i+2]}{Q^{\prime \prime }}=0,\;m=i+1,
\label{sp3.ap20}
\end{equation}
while the last equations in (\ref{sp3.ap15}--\ref{sp3.ap16}) are expressed
by 
\begin{equation}
\delta _{2}\stackrel{[i,0,k+j+1]}{R^{\prime }}=0,\;m=j,  \label{sp3.ap21}
\end{equation}
\begin{equation}
\delta _{1}\stackrel{[0,j,k+i+1]}{R^{\prime \prime }}=0,\;m=i.
\label{sp3.ap22}
\end{equation}
Their solutions are obviously given by 
\begin{equation}
\stackrel{\lbrack i-1,0,k+j+2]}{Q^{\prime }}=\delta _{2}\stackrel{%
[i-1,1,k+j+2]}{M^{\prime }},  \label{sp3.ap23}
\end{equation}
\begin{equation}
\stackrel{\lbrack 0,j-1,k+i+2]}{Q^{\prime \prime }}=\delta _{1}\stackrel{%
[1,j-1,k+i+2]}{M^{\prime \prime }},  \label{sp3.ap24}
\end{equation}
\begin{equation}
\stackrel{\lbrack i,0,k+j+1]}{R^{\prime }}=\delta _{2}\stackrel{[i,1,k+j+1]}{%
N^{\prime }},  \label{sp3.ap25}
\end{equation}
\begin{equation}
\stackrel{\lbrack 0,j,k+i+1]}{R^{\prime \prime }}=\delta _{1}\stackrel{%
[1,j,k+i+1]}{N^{\prime \prime }}.  \label{sp3.ap26}
\end{equation}
Introducing the solutions (\ref{sp3.ap23}) and (\ref{sp3.ap25}) in the
equation (\ref{sp3.ap17}) corresponding to $m=j$, respectively, the
solutions (\ref{sp3.ap24}) and (\ref{sp3.ap26}) in the equation (\ref
{sp3.ap18}) associated with $m=i$, we further infer the solutions to these
two equations like 
\begin{equation}
\stackrel{\lbrack i-1,1,k+j+1]}{Q^{\prime }}=\delta _{1}\stackrel{[i,1,k+j+1]%
}{N^{\prime }}+\delta _{2}\stackrel{[i-1,2,k+j+1]}{M^{\prime }}-\delta _{3}%
\stackrel{[i-1,1,k+j+2]}{M^{\prime }},  \label{sp3.ap27}
\end{equation}
\begin{equation}
\stackrel{\lbrack 1,j-1,k+i+1]}{Q^{\prime \prime }}=\delta _{2}\stackrel{%
[1,j,k+i+1]}{N^{\prime \prime }}+\delta _{1}\stackrel{[2,j-1,k+i+1]}{%
M^{\prime \prime }}-\delta _{3}\stackrel{[1,j-1,k+i+2]}{M^{\prime \prime }}.
\label{sp3.ap28}
\end{equation}
In order to solve the equations (\ref{sp3.ap15}) and (\ref{sp3.ap16}) for $%
m=j-1$, respectively, $m=i-1$, we employ one more time the solutions (\ref
{sp3.ap25}), respectively, (\ref{sp3.ap26}), which allows us to determine 
\begin{equation}
\stackrel{\lbrack i,1,k+j]}{R^{\prime }}=\delta _{2}\stackrel{[i,2,k+j]}{%
N^{\prime }}-\delta _{3}\stackrel{[i,1,k+j+1]}{N^{\prime }},
\label{sp3.ap29}
\end{equation}
\begin{equation}
\stackrel{\lbrack 1,j,k+i]}{R^{\prime \prime }}=\delta _{1}\stackrel{%
[2,j,k+i]}{N^{\prime \prime }}-\delta _{3}\stackrel{[1,j,k+i+1]}{N^{\prime
\prime }}.  \label{sp3.ap30}
\end{equation}
Using the results (\ref{sp3.ap27}) and (\ref{sp3.ap29}) in the equation (\ref
{sp3.ap17}) for $m=j-1$, respectively, (\ref{sp3.ap28}) and (\ref{sp3.ap30})
in the equation (\ref{sp3.ap18}) for $m=i-1$, we reach their solutions in
the form 
\begin{equation}
\stackrel{\lbrack i-1,2,k+j]}{Q^{\prime }}=\delta _{1}\stackrel{[i,2,k+j]}{%
N^{\prime }}+\delta _{2}\stackrel{[i-1,3,k+j]}{M^{\prime }}-\delta _{3}%
\stackrel{[i-1,2,k+j+1]}{M^{\prime }},  \label{sp3.ap31}
\end{equation}
\begin{equation}
\stackrel{\lbrack 2,j-1,k+i]}{Q^{\prime \prime }}=\delta _{2}\stackrel{%
[2,j,k+i]}{N^{\prime \prime }}+\delta _{1}\stackrel{[3,j-1,k+i]}{M^{\prime
\prime }}-\delta _{3}\stackrel{[2,j-1,k+i+1]}{M^{\prime \prime }}.
\label{sp3.ap32}
\end{equation}
Reprising the same operations, namely, inserting step by step the solutions
into the equations (\ref{sp3.ap15}--\ref{sp3.ap18}) with decreasing $m$, we
find the following solutions to the equations (\ref{sp3.ap11}--\ref{sp3.ap14}%
) 
\begin{equation}
\stackrel{\lbrack i,j-2,k+3]}{R^{\prime }}=\delta _{2}\stackrel{[i,j-1,k+3]}{%
N^{\prime }}-\delta _{3}\stackrel{[i,j-2,k+4]}{N^{\prime }},
\label{sp3.ap33}
\end{equation}
\begin{equation}
\stackrel{\lbrack i-2,j,k+3]}{R^{\prime \prime }}=\delta _{1}\stackrel{%
[i-1,j,k+3]}{N^{\prime \prime }}-\delta _{3}\stackrel{[i-2,j,k+4]}{N^{\prime
\prime }},  \label{sp3.ap34}
\end{equation}
\begin{equation}
\stackrel{\lbrack i-1,j-1,k+3]}{Q}=\delta _{1}\stackrel{[i,j-1,k+3]}{%
N^{\prime }}+\delta _{2}\stackrel{[i-1,j,k+3]}{M^{\prime }}-\delta _{3}%
\stackrel{[i-1,j-1,k+4]}{M^{\prime }},  \label{sp3.ap35}
\end{equation}
\begin{equation}
\stackrel{\lbrack i-1,j-1,k+3]}{Q}=\delta _{2}\stackrel{[i-1,j,k+3]}{%
N^{\prime \prime }}+\delta _{1}\stackrel{[i,j-1,k+3]}{M^{\prime \prime }}%
-\delta _{3}\stackrel{[i-1,j-1,k+4]}{M^{\prime \prime }}.  \label{sp3.ap36}
\end{equation}
At this stage, we notice that there appears a first restriction due to (\ref
{sp3.ap35}--\ref{sp3.ap36}), namely, 
\begin{eqnarray}
&&\delta _{1}\stackrel{[i,j-1,k+3]}{N^{\prime }}+\delta _{2}\stackrel{%
[i-1,j,k+3]}{M^{\prime }}-\delta _{3}\stackrel{[i-1,j-1,k+4]}{M^{\prime }}= 
\nonumber \\
&&\delta _{2}\stackrel{[i-1,j,k+3]}{N^{\prime \prime }}+\delta _{1}\stackrel{%
[i,j-1,k+3]}{M^{\prime \prime }}-\delta _{3}\stackrel{[i-1,j-1,k+4]}{%
M^{\prime \prime }}.  \label{sp3.ap37}
\end{eqnarray}
In the meantime, the solution to the equations (\ref{sp3.ap8}--\ref{sp3.ap9}%
) is expressed by 
\begin{equation}
\stackrel{\lbrack i,j-1,k+2]}{R^{\prime }}=\delta _{2}\stackrel{[i,j,k+2]}{%
N^{\prime }}-\delta _{3}\stackrel{[i,j-1,k+3]}{N^{\prime }},
\label{sp3.ap38}
\end{equation}
\begin{equation}
\stackrel{\lbrack i-1,j,k+2]}{R^{\prime \prime }}=\delta _{1}\stackrel{%
[i,j,k+2]}{N^{\prime \prime }}-\delta _{3}\stackrel{[i-1,j,k+3]}{N^{\prime
\prime }}.  \label{sp3.ap39}
\end{equation}
Looking now at the equation (\ref{sp3.ap10}), it only gives some new
restrictions as it should be satisfied for every of the solutions (\ref
{sp3.ap35}) and (\ref{sp3.ap36}). Thus if we use the results given by (\ref
{sp3.ap38}--\ref{sp3.ap39}) in the equation (\ref{sp3.ap10}), where we
replace $\stackrel{[i-1,j-1,k+3]}{Q}$ alternatively with the solutions (\ref
{sp3.ap35}) and (\ref{sp3.ap36}), we arrive at the new restrictions 
\begin{eqnarray}
&&\delta _{1}\delta _{2}\stackrel{[i,j,k+2]}{N^{\prime }}-\delta _{1}\delta
_{3}\stackrel{[i,j-1,k+3]}{N^{\prime }}+\delta _{2}\delta _{1}\stackrel{%
[i,j,k+2]}{N^{\prime \prime }}-\delta _{2}\delta _{3}\stackrel{[i-1,j,k+3]}{%
N^{\prime \prime }}=  \nonumber \\
&&\delta _{3}\delta _{1}\stackrel{[i,j-1,k+3]}{N^{\prime }}+\delta
_{3}\delta _{2}\stackrel{[i-1,j,k+3]}{M^{\prime }},  \label{sp3.ap40}
\end{eqnarray}
\begin{eqnarray}
&&\delta _{1}\delta _{2}\stackrel{[i,j,k+2]}{N^{\prime }}-\delta _{1}\delta
_{3}\stackrel{[i,j-1,k+3]}{N^{\prime }}+\delta _{2}\delta _{1}\stackrel{%
[i,j,k+2]}{N^{\prime \prime }}-\delta _{2}\delta _{3}\stackrel{[i-1,j,k+3]}{%
N^{\prime \prime }}=  \nonumber \\
&&\delta _{3}\delta _{2}\stackrel{[i-1,j,k+3]}{N^{\prime \prime }}+\delta
_{3}\delta _{1}\stackrel{[i,j-1,k+3]}{M^{\prime \prime }}.  \label{sp3.ap41}
\end{eqnarray}
The restrictions (\ref{sp3.ap37}) and (\ref{sp3.ap40}--\ref{sp3.ap41}) are
simultaneously satisfied if and only if there exist some elements $X^{\prime
}$, $X^{\prime \prime }$, $Y^{\prime }$, $Y^{\prime \prime }$, $Z^{\prime }$%
, $Z^{\prime \prime }$ and $W$ such that 
\begin{equation}
\stackrel{\lbrack i,j,k+2]}{N^{\prime }}=\stackrel{[i,j,k+2]}{N^{\prime
\prime }}+\delta _{1}\stackrel{[i+1,j,k+2]}{X^{\prime }}+\delta _{2}%
\stackrel{[i,j+1,k+2]}{X^{\prime \prime }}+\delta _{3}\stackrel{[i,j,k+3]}{W}%
,  \label{sp3.ap42}
\end{equation}
\begin{equation}
\stackrel{\lbrack i-1,j,k+3]}{N^{\prime \prime }}=\stackrel{[i-1,j,k+3]}{%
M^{\prime }}+\delta _{2}\stackrel{[i-1,j+1,k+3]}{Y^{\prime }}+\delta _{3}%
\stackrel{[i-1,j,k+4]}{Y^{\prime \prime }}+\delta _{1}\stackrel{[i,j,k+3]}{W}%
,  \label{sp3.ap43}
\end{equation}
\begin{equation}
\stackrel{\lbrack i,j-1,k+3]}{N^{\prime }}=\stackrel{[i,j-1,k+3]}{M^{\prime
\prime }}+\delta _{1}\stackrel{[i+1,j-1,k+3]}{Z^{\prime }}+\delta _{3}%
\stackrel{[i,j-1,k+4]}{Z^{\prime \prime }}-\delta _{2}\stackrel{[i,j,k+3]}{W}%
,  \label{sp3.ap44}
\end{equation}
where, in addition, there must exist an object $T$ that correlates $%
Y^{\prime \prime }$ and $Z^{\prime \prime }$ with $\stackrel{[i-1,j-1,k+4]}{%
M^{\prime }}$ and $\stackrel{[i-1,j-1,k+4]}{M^{\prime \prime }}$ involved
with the right hand-sides of the solutions (\ref{sp3.ap35}--\ref{sp3.ap36})
via the relations 
\begin{equation}
-\delta _{1}\stackrel{[i,j-1,k+4]}{Z^{\prime \prime }}+\delta _{2}\stackrel{%
[i-1,j,k+4]}{Y^{\prime \prime }}=\stackrel{[i-1,j-1,k+4]}{M^{\prime }}-%
\stackrel{[i-1,j-1,k+4]}{M^{\prime \prime }}+\delta _{3}\stackrel{%
[i-1,j-1,k+5]}{T}.  \label{sp3.ap45}
\end{equation}
Thus, we have completely elucidated the solutions to the equations (\ref
{sp3.ap8}--\ref{sp3.ap10}). We are now left only with the equations (\ref
{sp3.ap6}--\ref{sp3.ap7}), which are obviously solved by 
\begin{equation}
\stackrel{\lbrack i,j,k+1]}{R}=\delta _{2}\stackrel{[i,j+1,k+1]}{N^{\prime }}%
-\delta _{3}\stackrel{[i,j,k+2]}{N^{\prime }},  \label{sp3.ap46}
\end{equation}
\begin{equation}
\stackrel{\lbrack i,j,k+1]}{R}=\delta _{1}\stackrel{[i+1,j,k+1]}{N^{\prime
\prime }}-\delta _{3}\stackrel{[i,j,k+2]}{N^{\prime \prime }},
\label{sp3.ap47}
\end{equation}
where $\stackrel{[i,j,k+2]}{N^{\prime }}$ and $\stackrel{[i,j,k+2]}{%
N^{\prime \prime }}$ are linked through the relation (\ref{sp3.ap42}). On
the other hand, the last two formulas imply a new restriction, namely 
\begin{equation}
\delta _{2}\stackrel{[i,j+1,k+1]}{N^{\prime }}-\delta _{3}\stackrel{[i,j,k+2]%
}{N^{\prime }}=\delta _{1}\stackrel{[i+1,j,k+1]}{N^{\prime \prime }}-\delta
_{3}\stackrel{[i,j,k+2]}{N^{\prime \prime }}.  \label{sp3.ap48}
\end{equation}
Using now (\ref{sp3.ap42}) in (\ref{sp3.ap48}), it follows that the last
restriction is equivalent to 
\begin{equation}
\delta _{1}\left( \stackrel{[i+1,j,k+1]}{N^{\prime \prime }}-\delta _{3}%
\stackrel{[i+1,j,k+2]}{X^{\prime }}\right) =\delta _{2}\left( \stackrel{%
[i,j+1,k+1]}{N^{\prime }}+\delta _{3}\stackrel{[i,j+1,k+2]}{X^{\prime \prime
}}\right) ,  \label{sp3.ap49}
\end{equation}
which is satisfied if and only if there exist some elements $F$, $G$ and $H$
such that 
\begin{equation}
\stackrel{\lbrack i+1,j,k+1]}{N^{\prime \prime }}-\delta _{3}\stackrel{%
[i+1,j,k+2]}{X^{\prime }}=-\delta _{2}\stackrel{[i+1,j+1,k+1]}{F}+\delta _{1}%
\stackrel{[i+2,j,k+1]}{G},  \label{sp3.ap50}
\end{equation}
\begin{equation}
\stackrel{\lbrack i,j+1,k+1]}{N^{\prime }}+\delta _{3}\stackrel{[i,j+1,k+2]}{%
X^{\prime \prime }}=\delta _{1}\stackrel{[i+1,j+1,k+1]}{F}+\delta _{2}%
\stackrel{[i,j+2,k+1]}{H}.  \label{sp3.ap51}
\end{equation}
Finally, substituting any of the solutions (\ref{sp3.ap46}--\ref{sp3.ap47})
(which now coincide) into the equation (\ref{sp3.ap5}) and taking into
account (\ref{sp3.ap50}--\ref{sp3.ap51}), we deduce that its solution is
provided by 
\begin{equation}
\stackrel{\lbrack i,j,k]}{F}=\delta _{3}\delta _{2}\delta _{1}\stackrel{%
[i+1,j+1,k+1]}{F},  \label{sp3.ap52}
\end{equation}
which proves the theorem.

\begin{theorem}
Let $\stackrel{[m]}{F}\in \mathcal{A}^{\prime }$, with $\mathrm{res}\left( 
\stackrel{[m]}{F}\right) =m>0$, be such that 
\begin{equation}
\stackrel{\lbrack m]}{F}=\sum\limits_{p+q+r=m}\stackrel{[p,q,r]}{F}.
\label{sp3.ap53}
\end{equation}
Assume that: 1. $\delta \stackrel{[m]}{F}=0$, and 2. in the sum (\ref
{sp3.ap53}) only terms with $p\leq i$, $q\leq j$, $r\leq k$ occur, for some $%
i$, $j$, $k$, such that $i+j+k>m$ (strictly). Then, we can represent $%
\stackrel{[m]}{F}$ under the form 
\begin{equation}
\stackrel{\lbrack m]}{F}=\delta \stackrel{[m+1]}{P},  \label{sp3.ap54}
\end{equation}
where 
\begin{equation}
\stackrel{\lbrack m+1]}{P}=\sum\limits_{\bar{p}+\bar{q}+\bar{r}=m+1}%
\stackrel{[\bar{p},\bar{q},\bar{r}]}{P},  \label{sp3.ap55}
\end{equation}
involves only terms with $\bar{p}\leq i$, $\bar{q}\leq j$ and $\bar{r}\leq k$%
.
\end{theorem}

{\textbf{Proof.}} According to the assumption 2, it follows that 
\begin{eqnarray}
&&\stackrel{[m]}{F}=\stackrel{[i,j,m-i-j]}{F}+\stackrel{[i-1,j,m-i-j+1]}{F}+%
\stackrel{[i,j-1,m-i-j+1]}{F}+  \nonumber \\
&&\cdots +\stackrel{[i,m-i-k,k]}{F}+\cdots +\stackrel{[m-j-k,j,k]}{F},
\label{sp3.ap56}
\end{eqnarray}
being understood that $\stackrel{[a,b,c]}{F}=0$ if $a<0$, or $b<0$, or $c<0$%
. It is clear that $\stackrel{[i,j,m-i-j]}{F}$ is the term in $F$ with the
lowest third component of the resolution tridegree (if $m-i-j<0$, then this
piece vanishes). Then, from $\delta \stackrel{[m]}{F}=0$, it follows that $%
\delta _{3}\stackrel{[i,j,m-i-j]}{F}=0$, such that, using $H_{i^{\prime
},j^{\prime },k^{\prime }}\left( \delta _{3}\right) =0$, for $i^{\prime
}+j^{\prime }+k^{\prime }>0$, we deduce that 
\begin{equation}
\stackrel{\lbrack i,j,m-i-j]}{F}=\delta _{3}\stackrel{[i,j,m-i-j+1]}{%
P^{\prime }},  \label{sp3.ap57}
\end{equation}
where $\stackrel{[i,j,m-i-j+1]}{P^{\prime }}\equiv 0$ if $m-i-j<0$. In the
meantime, we have that $m-i-j+1\leq k$ because $m<i+j+k$. If we subtract $%
\delta \stackrel{[i,j,m-i-j+1]}{P^{\prime }}$ from $\stackrel{[m]}{F}$, we
infer that 
\begin{eqnarray}
&&\stackrel{[m]}{F}-\delta \stackrel{[i,j,m-i-j+1]}{P^{\prime }}=\stackrel{%
[i-1,j,m-i-j+1]}{F^{\prime }}+\stackrel{[i,j-1,m-i-j+1]}{F^{\prime }}+\cdots
\nonumber \\
&&+\stackrel{[i,m-i-k,k]}{F}+\cdots +\stackrel{[m-j-k,j,k]}{F}.
\label{sp3.ap58}
\end{eqnarray}
Next, we act similarly and remove the terms of lowest third component of the
resolution tridegree in the right hand-side of (\ref{sp3.ap58}), namely, $%
\stackrel{[i-1,j,m-i-j+1]}{F^{\prime }}+\stackrel{[i,j-1,m-i-j+1]}{F^{\prime
}}$. If we proceed in the same manner, we reach the last step 
\begin{equation}
\stackrel{\lbrack m]}{F}-\delta \tilde{P}=\stackrel{[m-j-k,j,k]}{F^{\prime }}%
,  \label{sp3.ap59}
\end{equation}
such that $\delta \stackrel{[m]}{F}=0$ becomes equivalent to 
\begin{eqnarray}
&&\left( \delta \stackrel{[m-j-k,j,k]}{F^{\prime }}=0\right) \Leftrightarrow
\nonumber \\
&&\left( \delta _{1}\stackrel{[m-j-k,j,k]}{F^{\prime }}=0,\delta _{2}%
\stackrel{[m-j-k,j,k]}{F^{\prime }}=0,\delta _{3}\stackrel{[m-j-k,j,k]}{%
F^{\prime }}=0\right) .  \label{sp3.ap60}
\end{eqnarray}
On the other hand, the previous theorem ensures that 
\begin{eqnarray}
&&\stackrel{[m-j-k,j,k]}{F^{\prime }}=\delta _{1}\delta _{2}\delta _{3}%
\stackrel{[m-j-k+1,j+1,k+1]}{Q}=  \nonumber \\
&&\left( \delta _{1}+\delta _{2}+\delta _{3}\right) \delta _{2}\delta _{3}%
\stackrel{[m-j-k+1,j+1,k+1]}{Q},  \label{sp3.ap61}
\end{eqnarray}
due to the nilpotency and anticommutativity of $\delta _{2}$ and $\delta
_{3} $, such that we can further write 
\begin{equation}
\stackrel{\lbrack m-j-k,j,k]}{F^{\prime }}=\delta \stackrel{[m-j-k+1,j,k]}{%
P^{\prime }},  \label{sp3.ap62}
\end{equation}
with $\stackrel{[m-j-k+1,j,k]}{P^{\prime }}=\delta _{2}\delta _{3}\stackrel{%
[m-j-k+1,j+1,k+1]}{Q}$. Then, we find that 
\begin{equation}
\stackrel{\lbrack m]}{F}=\delta \left( \stackrel{[m+1]}{P}\right) ,
\label{sp3.ap63}
\end{equation}
where 
\begin{equation}
\stackrel{\lbrack m+1]}{P}=\stackrel{[i,j,m-i-j+1]}{P^{\prime }}+\cdots +%
\stackrel{[i,m-i-k+1,k]}{P^{\prime }}+\cdots +\stackrel{[m-j-k+1,j,k]}{%
P^{\prime }}.  \label{sp3.ap64}
\end{equation}
It is now easy to see that $m-i-j+1\leq k$, $m-i-k+1\leq j$ and $m-j-k+1\leq
i$, as we supposed that $m<i+j+k$, hence $\stackrel{[m+1]}{P}$ includes only
terms of resolution tridegrees $\left( \bar{p},\bar{q},\bar{r}\right) $,
with $\bar{p}\leq i$, $\bar{q}\leq j$ and $\bar{r}\leq k$, such that $\bar{p}%
+\bar{q}+\bar{r}=m+1$. This ends the proof.

\end{document}